\documentclass[12pt]{article}

\usepackage{array} 
\usepackage{amssymb}
\usepackage{graphics,graphpap}
\usepackage{graphicx}
\usepackage{color}
\usepackage{graphicx}
\usepackage{dcolumn}
\usepackage{epsfig}
\usepackage{epstopdf}
\usepackage{bm}
\usepackage{amsmath}
\usepackage{amsfonts}
\usepackage{textcomp}
\usepackage{setspace}

\usepackage{multirow}
\usepackage{longtable}
\usepackage{booktabs}
\usepackage{cellspace}
\usepackage{bigdelim}
\usepackage{bigstrut}
\usepackage{subfig}

\newcommand{\beq}{\begin{eqnarray}}
\newcommand{\eeq}{\end{eqnarray}}

\newcommand{\bmp}{\noindent\begin{minipage}{16cm}}
\newcommand{\emp}{\end{minipage}\vskip 7mm} 

\def\drawbox#1#2{\hrule height#2pt
        \hbox{\vrule width#2pt height#1pt \kern#1pt
              \vrule width#2pt}
              \hrule height#2pt}

\def\Asym#1#2{\vcenter{\vbox{\drawbox{#1}{#2}
              \kern-#2pt 
              \drawbox{#1}{#2}}}}

\def\simge{\mathrel{%
   \rlap{\raise 0.511ex \hbox{$>$}}{\lower 0.511ex \hbox{$\sim$}}}}

\def\simle{\mathrel{
   \rlap{\raise 0.511ex \hbox{$<$}}{\lower 0.511ex \hbox{$\sim$}}}}

\def\s#1{\setbox0=\hbox{$#1$}%
\rlap{\ifdim\wd0>.7em\kern.22\wd0\else\kern.1\wd0\fi /}#1}

\newcommand{\hnut}{h_\nu^{\scriptstyle T}}

\newcommand{\keffi}[1] {\overset{\hspace{-0.8cm} \scriptscriptstyle (#1)}{\kappa_{\scriptscriptstyle(11),ij}}}
\newcommand{\keffid}[1] {\overset{\hspace{-0.8cm} \scriptscriptstyle (#1)}{\kappa_{\scriptscriptstyle(22),ij}}}
\newcommand{\suph}[1]{\overset{\scriptscriptstyle{(#1)}}{h}_\nu}
\newcommand{\supht}[1]{\overset{\scriptscriptstyle{(#1)}}{\hnut}}
\newcommand{\smas}[1]{\overset{\scriptscriptstyle{(#1)}}{M}}

\begin{document}

\begin{titlepage}
\title{\vspace*{-2.0cm}
\bf\Large
Running of Radiative Neutrino Masses:\\ The Scotogenic Model
\\[5mm]\ }

\author{
Romain Bouchand\thanks{email: \tt bouchand@kth.se}\ \ \ and\ \ 
Alexander Merle\thanks{email: \tt amerle@kth.se}\
\\ \\
{\normalsize \it Department of Theoretical Physics, School of Engineering Sciences,}\\
{\normalsize \it KTH Royal Institute of Technology -- AlbaNova University Center,}\\
{\normalsize \it Roslagstullsbacken 21, 106 91 Stockholm, Sweden}
}
\date{\today}
\maketitle
\thispagestyle{empty}

\begin{abstract}
\noindent
We study the renormalization group equations of Ma's scotogenic model, which generates an active neutrino mass at 1-loop level. In addition to other benefits, the main advantage of the mechanism exploited in this model is to lead to a natural loop-suppression of the neutrino mass, and therefore to an explanation for its smallness. However, since the structure of the neutrino mass matrix is altered compared to the ordinary type~I seesaw case, the corresponding running is altered as well. We have derived the full set of renormalization group equations for the scotogenic model which, to our knowledge, had not been presented previously in the literature. This set of equations reflects some interesting structural properties of the model, and it is an illustrative example for how the running of neutrino parameters in radiative models is modified compared to models with tree-level mass generation. We also study a simplified numerical example to illustrate some general tendencies of the running. Interestingly, the structure of the RGEs can be exploited such that a bimaximal leptonic mixing pattern at the high-energy scale is translated into a valid mixing pattern at low energies, featuring a large value of $\theta_{13}$. This suggests very interesting connections to flavour symmetries.
\end{abstract}
\end{titlepage}

\section{\label{sec:intro}Introduction}

The leptonic sector of the Standard Model (SM) of particle physics has undergone several modifications within the last decades. Not only have we learned from oscillation experiments that neutrinos must have a non-zero mass, but we have meanwhile also measured the corresponding mixing angles~\cite{Schwetz:2011qt,Schwetz:2011zk}. This includes in particular the very recent spectacular measurements of the previously unknown mixing angle $\theta_{13}$ by the Daya Bay~\cite{An:2012eh} and RENO~\cite{Ahn:2012nd} collaborations, after hints from MINOS~\cite{Adamson:2011qu}, T2K~\cite{Abe:2011sj}, and Double Chooz~\cite{Abe:2011fz}. What is still missing, however, is the knowledge of the absolute neutrino mass scale: Although we have limits from kinematical measurements such as Troitsk~\cite{Lobashev:2000vb} and Mainz~\cite{Kraus:2004zw}, from experiments on neutrino-less double $\beta$ decay~\cite{KlapdorKleingrothaus:2000sn,Andreotti:2010vj}, and from cosmology~\cite{Komatsu:2010fb}, the only decent statement we can make is that the true neutrino mass scale is somewhere below $1$~eV.

\begin{figure}[t]
\centering
\includegraphics[width=7.5cm]{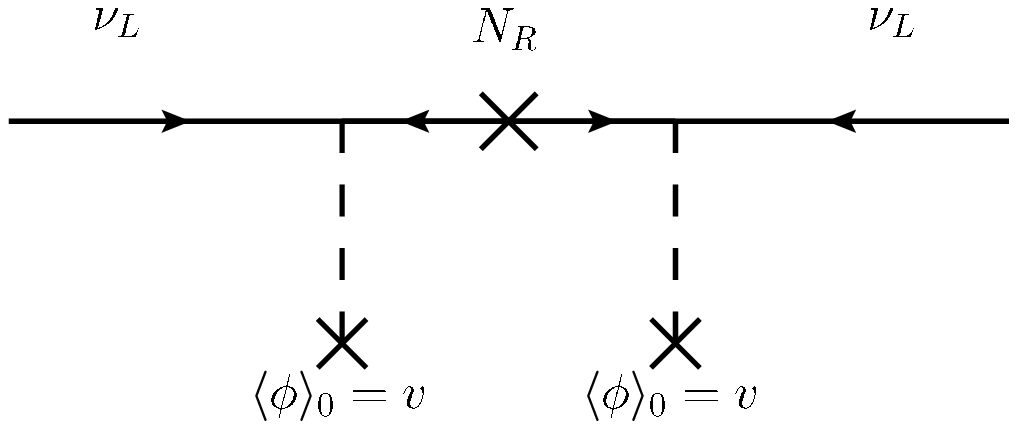}
\includegraphics[width=6.5cm]{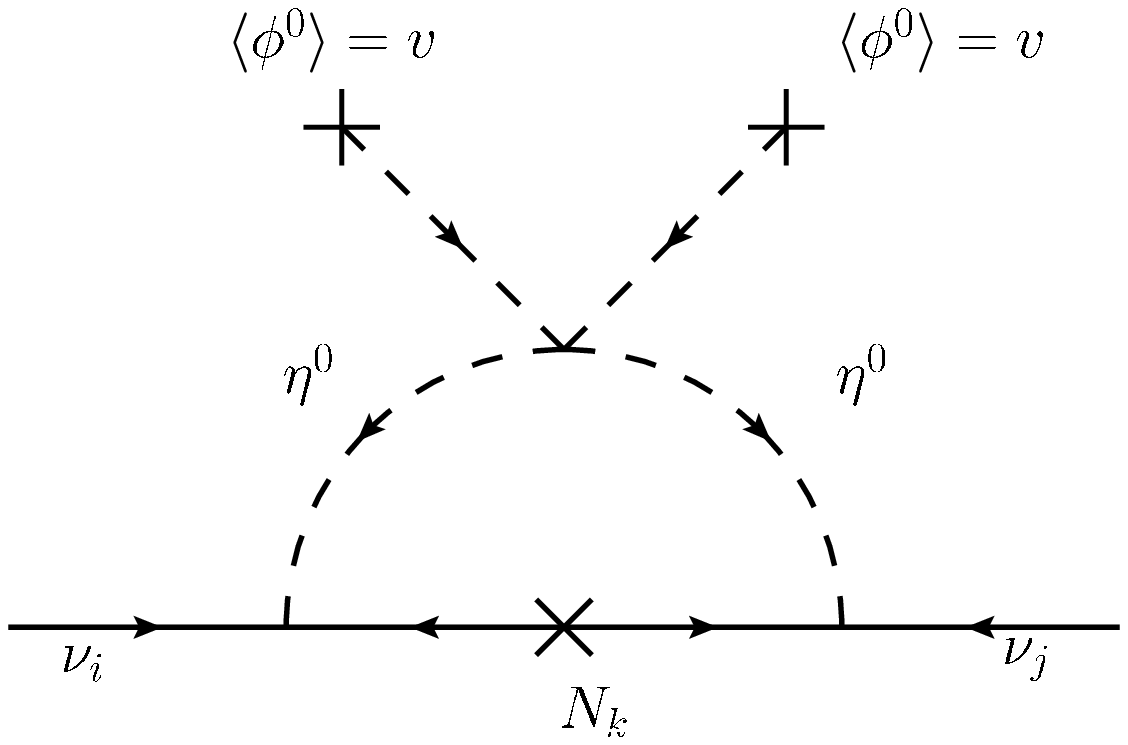}
\caption{\label{fig:seesaws} Tree-level and radiative seesaw mechanisms.}
\end{figure}

Nevertheless, generating a mass scale $\lesssim 1$~eV in a model is far from trivial: In the classic version of the SM, neutrinos are strictly massless, but even if we extend the SM by right-handed neutrinos to generate Dirac neutrino mass terms, the tininess of the neutrino mass scale compared to the scale of electroweak symmetry breaking (EWSB) looks highly unnatural. Probably the most frequently discussed solution to this problem is to introduce a Majorana mass term for the right-handed neutrino states and to exploit the potential large size of the corresponding masses through the famous seesaw (type~I) mechanism~\cite{Minkowski:1977sc,Yanagida:1979as,GellMann:1980vs,Glashow:1979nm,Mohapatra:1979ia}, as depicted on the left panel of Fig.~\ref{fig:seesaws}. Seesaw-type settings have also been investigated in the context of renormalization group equations (RGEs)~\cite{Antusch:2005gp,Antusch:2002rr,kerstendipl,KerstenPhD,Schmidt:2007nq,Bergstrom:2010id}, in order to get a more detailed picture of the change of neutrino masses and mixing parameters with the energy scale, usually referred to as \emph{RG-running} or simply  \emph{running}.

An alternative and very attractive way to explain the smallness of neutrino masses is to assume that they are strictly zero at tree-level, but receive non-zero higher order corrections. Maybe the most simple such framework can be found in Ma's \emph{scotogenic model}~\cite{Ma:2006km} (for simplicity called ``Ma-model'' in this paper), in which the introduction of only a few new fields and one additional symmetry compared to the SM leads to a non-zero neutrino mass at 1-loop level, cf.\ right panel of Fig.~\ref{fig:seesaws}. In fact, this diagram exhibits a structure that is somewhat reminiscent of a seesaw-type situation, which is why sometimes the corresponding suppression is dubbed as \emph{radiative seesaw}.

The main idea that led to this paper is to investigate the running of neutrino parameters in models with a radiative neutrino mass, as the rich structure of the corresponding diagrams is likely to introduce interesting and yet unknown features in the running, which are not present in tree-level realizations of the seesaw mechanism. To our knowledge, there exists no such study in the literature with the focus put on the neutrino sector in radiative models, and we aim to start this enterprise by a study devoted to the RGEs of the Ma-model. Naturally, this could be extended to other radiative models for neutrino masses, such as the Zee-Babu model~\cite{Babu:1988ki,Zee:1985id} or the Aoki-Kanemura-Seto model~\cite{Aoki:2008av,Aoki:2011zg}. In particular the interplay between the scalar and the lepton sectors has the potential to reveal interesting new effects, as we will already see in this study.

However, we want to stress that several studies are already available which investigate e.g.\ limiting cases of our framework or subsets (or generalizations of subsets) of certain sectors of the Ma-model. A particular example for such a case would be the investigations of the RGEs of a general Two Higgs Doublet Model (THDM). Whenever applicable in this paper, we will refer to the corresponding works treating these related frameworks.

This paper is organized as follows: In Sec.~\ref{sec:Ma-model}, we review Ma's scotogenic model and discuss the different effective theories arising when subsequently integrating out the heavy neutrino fields. Next, in Sec.~\ref{sec:Ma-tching}, we discuss in detail the matching conditions at the boundaries between the respective theories, which in our case have to be consistently imposed at 1-loop level. Our main results, the explicit RGEs at 1-loop level are presented in Sec.~\ref{sec:RGEs}. After that, we present a numerical exemplifying study (in a slightly simplified framework) in Sec.~\ref{sec:Numerics}, in order to illustrate how to use our results. We finally conclude in Sec.~\ref{sec:Conclusions}.

\section{\label{sec:Ma-model}Ma's scotogenic model}

The so-called \emph{scotogenic model} has been discussed by Ma~\cite{Ma:2006km}, and in the following we will therefore call it \emph{Ma-model} for simplicity. In this section, we will first review this model, and then discuss some of its low-energy limits, which we will also use in our calculations later on.

\subsection{\label{sec:Ma-pure}The Ma-model as full theory}

The Ma-model is a very economic extension of the SM, which however exhibits a variety of very interesting features. Some of its nicest aspects are its possibilities for Dark Matter (DM) candidates, the radiative generation of neutrino masses, and the collider phenomenology of relatively light right-handed neutrinos. Compared to the SM, the model contains three new crucial ingredients:
\begin{enumerate}

\item In addition to the SM Higgs $\Phi$, there is a second scalar doublet $\eta \sim \left(\mathbf{2}, 1/2\right)$, which has the same quantum numbers under $SU(2)_L\times U(1)_Y$ as the ordinary Higgs.

\item Furthermore, there are three right-handed neutrinos $N_k \sim \left(\mathbf{1}, 0\right)$, which have a Majorana mass term, $\mathcal{L}=-\frac{1}{2} M_k \overline{N_k^c} N_k$.\footnote{Note that we have already used the freedom of rotating the basis such that this Majorana mass term is diagonal.}

\item Finally, there is an exact (discrete) $Z_2$ parity, under which all SM-fields transform trivially (i.e., they have $Z_2$-charge $+1$), while all new fields, $\eta$ and $N_k$, are odd (i.e., they have $Z_2$-charge $-1$). Because of the non-trivial charge under this symmetry, the new doublet $\eta$ is often dubbed as \emph{inert} or \emph{dark} scalar doublet.

\end{enumerate}

These few assumptions lead to a variety of very interesting consequences, some of which we will list in the following:

\begin{itemize}

\item Due to the $Z_2$-symmetry being unbroken, the lightest particle charged under this symmetry will be absolutely stable. If this particle is electrically neutral, i.e., it is one of the $N_k$ or one of the neutral components of the scalar field $\eta$, it would be a good candidate particle for DM~\cite{Kubo:2006yx,Ma:2006ms,LopezHonorez:2006gr,Sierra:2008wj,Gelmini:2009xd,Suematsu:2010gv}.

\item Since the standard neutrino Yukawa coupling, $\mathcal{L}_{\rm Yuk,\nu} = - \overline{L} Y_\nu \tilde \Phi N_R + h.c.$ where $\tilde \Phi = i \sigma_2 \Phi^*$, is forbidden by the exact $Z_2$ symmetry, this term cannot lead to a tree-level neutrino mass. There is an alternative Yukawa coupling involving the new scalar, $\mathcal{L}'_{\rm Yuk,\nu} = - \overline{L} h_\nu \tilde \eta N_R + h.c.$, but also this term does not generate a neutrino mass at tree-level. However, it does lead to a non-vanishing neutrino mass at 1-loop level, cf.\ right panel of Fig.~\ref{fig:seesaws}. The resulting formula for the neutrino mass matrix is:
\begin{equation}
\mathcal{M}_{\nu,ij} = \frac{h_{ik} h_{jk}}{16\pi^2} M_k \left[\frac{m_R^2}{m_R^2-M_k^2}\ln \frac{m_R^2}{M_k^2}-\frac{m_I^2}{m_I^2-M_k^2}\ln \frac{m_I^2}{M_k^2}\right].
 \label{eq:Ma-matrix}
\end{equation}
This expression involves the scalar masses $m_R$ ($m_I$) of the real (imaginary) parts of the electrically neutral component $\eta^0$ of the new scalar. These masses originate from the scalar potential:
\begin{equation}
\begin{split}
V_{\mathrm{scalar}}&=m_1^2 \Phi^\dagger \Phi + m_2^2 \eta^\dagger \eta + \frac{1}{2}  \lambda_1 \left( \Phi^\dagger \Phi \right)^2+\frac{1}{2} \lambda_2 \left(\eta^\dagger \eta \right)^2\\
&\ \ \ \ +\lambda_3 \left( \Phi^\dagger \Phi \right) \left( \eta^\dagger \eta \right) +\lambda_4 \left( \Phi^\dagger \eta \right) \left( \eta^\dagger \Phi \right) +\frac{1}{2}\lambda_5 \left[\left( \Phi^\dagger \eta \right)^2 + \mathrm{h.c.}\right],
\end{split}
\label{eq:Pot}
\end{equation}
where all parameters can be taken to be real, but we need to choose $m_1^2<0$ and $m_2^2>0$ in order to have a VEV of $\langle \Phi^0 \rangle = v = 174$~GeV only, while keeping $\langle \eta^0 \rangle = 0$. This choice leads to the following scalar masses (just as for a general THDM~\cite{Eriksson:2009ws} with $\lambda_{6,7}=0$), where $h$ is the SM-like Higgs particle:
\begin{eqnarray}
 m_h^2 &=& 2\lambda_1 v^2,\nonumber\\
 m_R^2 \equiv m^2(\eta_R^0) &=& m_2^2+(\lambda_3+\lambda_4+\lambda_5)v^2,\nonumber\\
 m_I^2 \equiv m^2(\eta_I^0) &=& m_2^2+(\lambda_3+\lambda_4-\lambda_5)v^2,\nonumber\\
 m^2(\eta^\pm) &=& m_2^2+\lambda_3 v^2.
 \label{eq:scalar_masses}
\end{eqnarray}

The formula for the light neutrino masses, Eq.~\eqref{eq:Ma-matrix}, exhibits a behaviour similar to that of a mass matrix resulting from a seesaw mechanism, since the Dirac Yukawa couplings are suppressed by some high mass scale. Hence this mechanism is often referred to as \emph{radiative seesaw}. However, due to the additional loop suppression factor $(16\pi^2)^{-1}$, and due to a hidden proportionality to the small coupling $\lambda_5$, only part of the suppression has to come from the masses of the heavy particles involved. Accordingly, the masses of the right-handed neutrinos can have relatively low values, i.e.\ around the TeV-scale, similar to the masses of the new scalar, thereby leading to potentially interesting collider signatures~\cite{Sierra:2008wj,Cao:2007rm,Atwood:2007zza,Haba:2011nb}.

\item Due to the extended scalar sector, this model typically leads to potentially observables rates of lepton flavour violating (LFV) processes~\cite{Kubo:2006yx,Sierra:2008wj}. In turn, when trying to explain a more detailed structure in the flavour sector, exactly these LFV processes tend to spoil the application of flavour symmetries to predict observables like mixing angles~\cite{Adulpravitchai:2009gi}.

\item On the other hand the extension of the Ma-model by an additional Left-Right symmetry can yield a completely new mechanism to explain large but not necessarily maximal mixing angles, the so-called \emph{radiative transmission of hierarchies}~\cite{Adulpravitchai:2009re}. This mechanism could be an alternative to flavour symmetries and in that way resolve the problem mentioned in the previous point.

\end{itemize}

As we can see, there are indeed many interesting aspects of this model discussed in the literature. What is, however, to our knowledge lacking up to now is a thorough treatment of the running of the neutrino mass matrix in the Ma-model. Since running is unavoidable, we should carefully check in how far the behaviour changes in radiative models, like the one under consideration, compared to models with tree-level neutrino masses. We could expect at least some qualitative changes, due to different diagrams and structures appearing in the calculation. Furthermore, since the neutrino mass is zero at tree-level, its 1-loop correction is actually the decisive term, and the running of this term could be potentially large. It is therefore important to carefully check if consistency of this model with low-energy data arises naturally, or if it maybe could be under some pressure due to the running effects being too strong.

\subsection{\label{sec:Ma-EFT}The effective field theories appearing}

In particle physics, it is well-known that a separation of scales translates into a decoupling of the heavy degrees of freedom at low energies, which goes under the name of the \emph{Appelquist-Carazzone} theorem~\cite{AppCar}, leading directly to so-called \emph{effective field theories} (EFTs)~\cite{Weinberg}. In practice, this mans that we can encode the effects of new physics at high energies into \emph{effective} operators, which contain only fields of the low-energy theory, but which have a mass dimension $d$ larger than four. In an EFT, we solely take into account the terms in the Lagrangian related to fields which are \emph{relevant} at the corresponding energy scale, i.e., we drop the terms involving the momenta of very heavy fields. It is this procedure that leads to the higher-dimensional terms. Note that there is an upper bound on the validity of any effective field theory, usually denoted as the cutoff $\Lambda$, inverse powers of which suppress the effective operators. Actually, EFTs are just a way to ``parametrize our lack of knowledge'' of the underlying low-distance physics.

We now take on a \emph{top-down} approach, in which we extract the EFT from a more fundamental high-energy theory. This means that we start with a completely renormalizable high-energy theory, run it down to lower energies by gradually decoupling the heavy fields, and eventually obtain the low-energy footprints of the high-energy regime of this theory. In order to do this we make use of  the corresponding \emph{renormalization group equations}. In the case of the Ma-Model, we will consider the decoupling of the heavy neutrinos -- they will be \emph{integrated out} -- and we will investigate which terms appear in the resulting low-energy effective Lagrangians. Note that we have assumed the inert scalar to be lighter than the right-handed neutrinos for the sake of an example. Then, in a manner very similar to a type~I seesaw mechanism, integrating out the heavy neutrinos will lead to $d=5$ operators coupling two lepton doublets to two scalar doublets.

\begin{figure}[t]
\centering
\includegraphics[width=3.9cm]{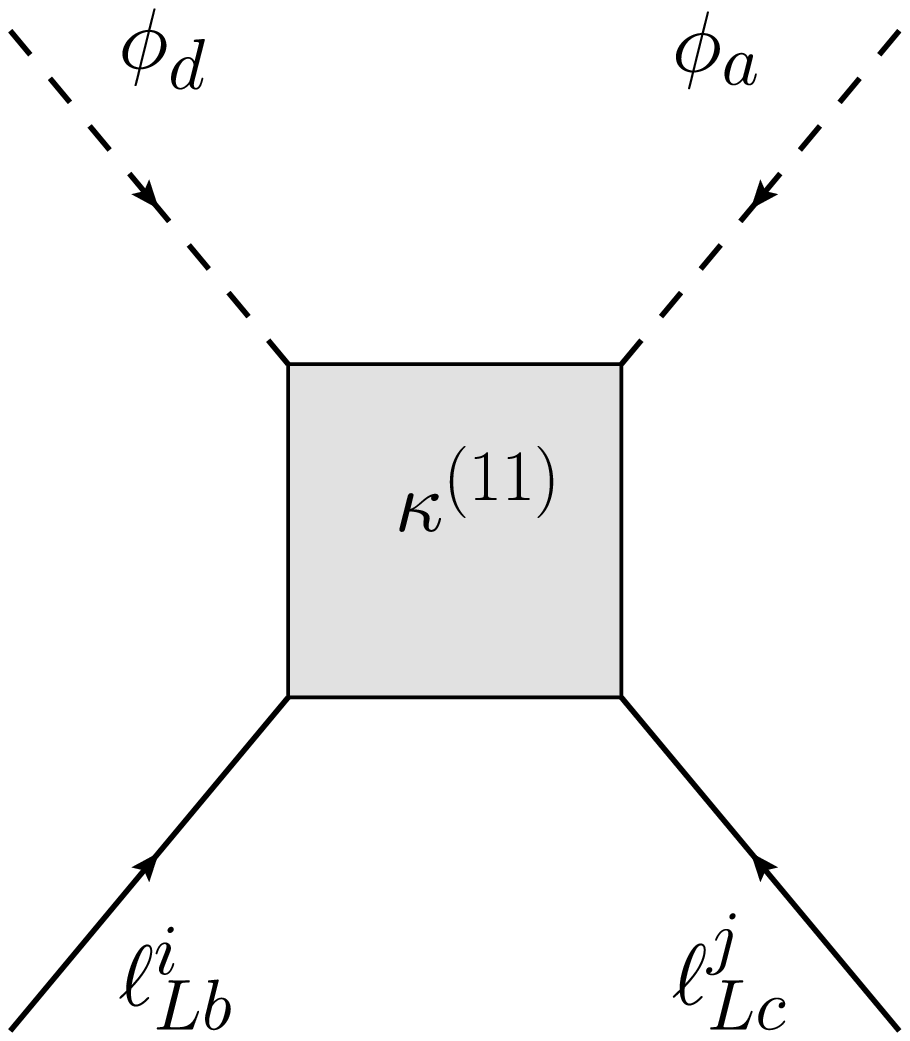}
\includegraphics[width=3.9cm]{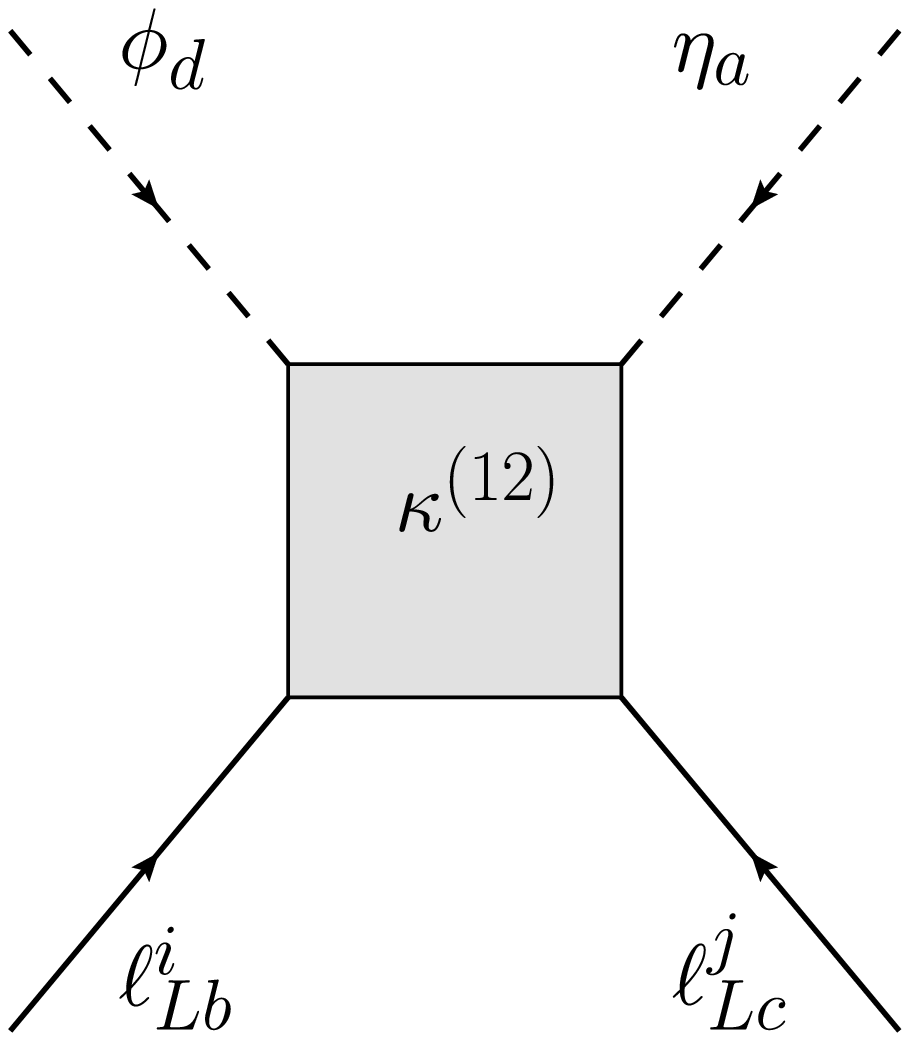}
\includegraphics[width=3.9cm]{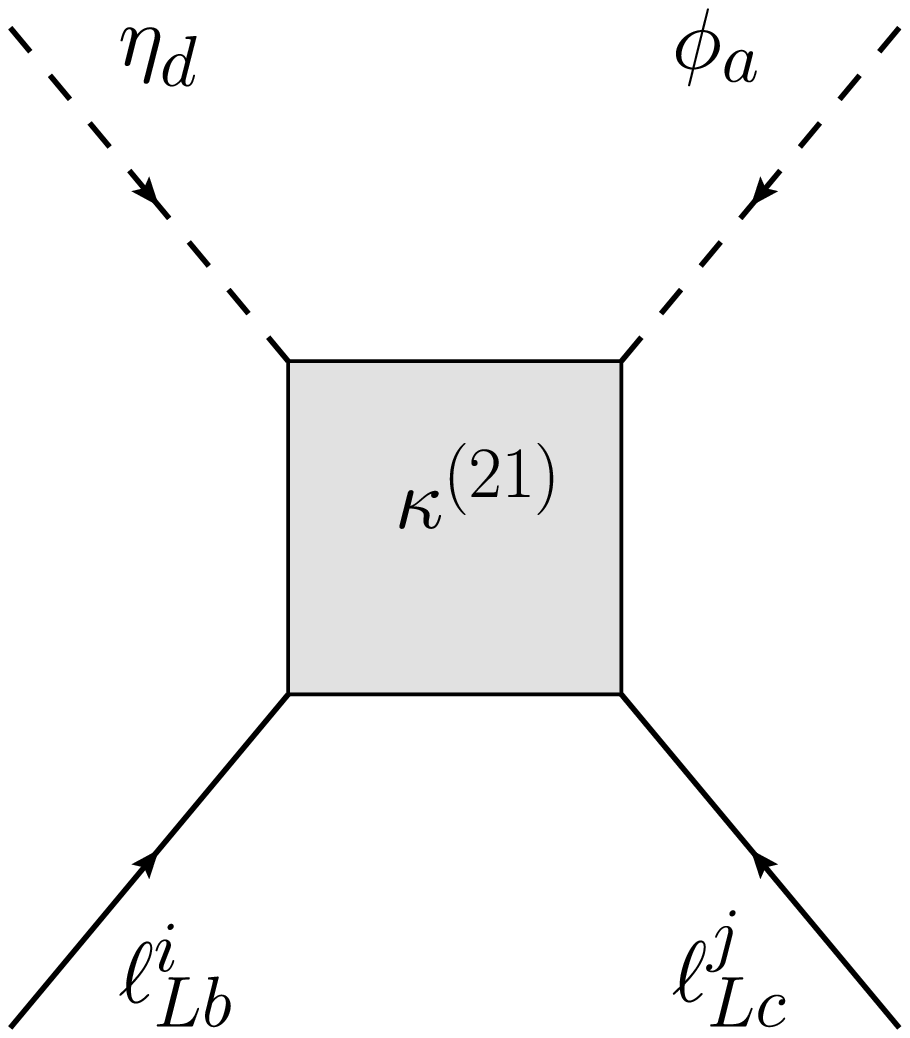}
\includegraphics[width=3.9cm]{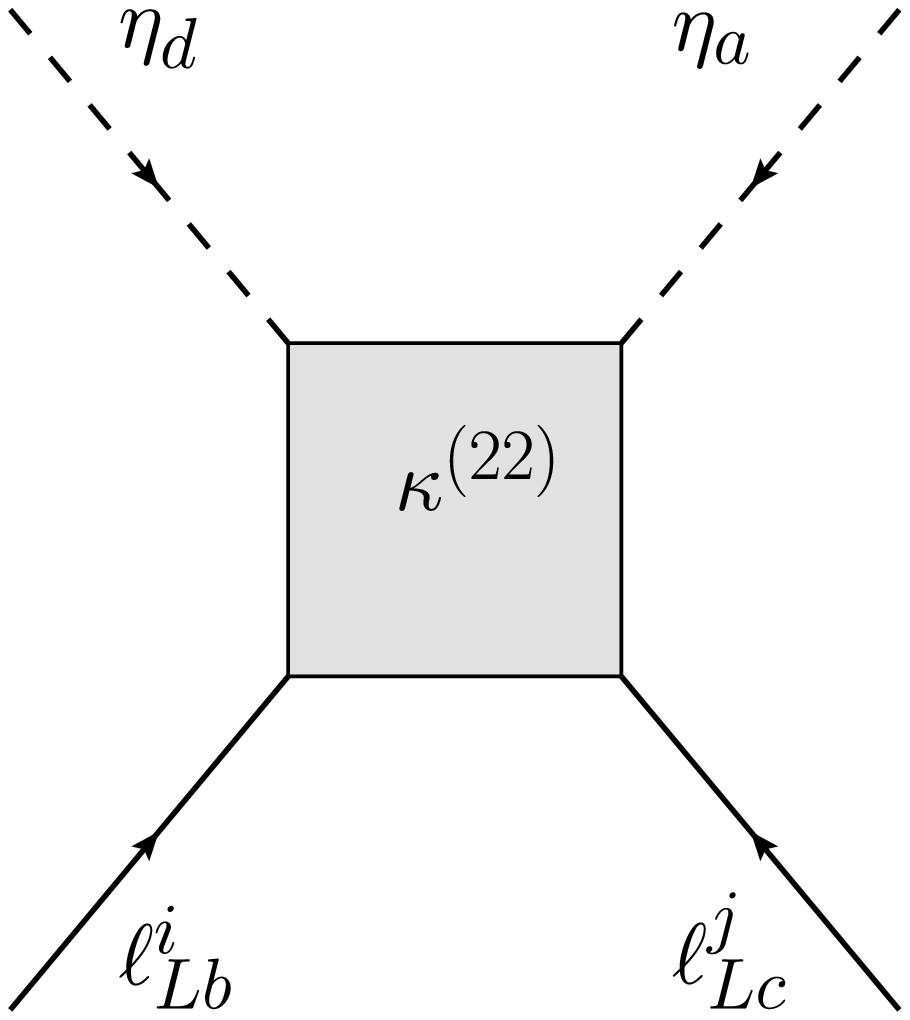}
\caption{\label{fig:eff} The potential Weinberg-like operators in a general THDM. In the Ma-model, the two operators $\kappa^{(12)}$ and $\kappa^{(21)}$ are forbidden by the unbroken $Z_2$ symmetry.}
\end{figure}

\begin{figure}[t]
\centering
\includegraphics[width=7cm]{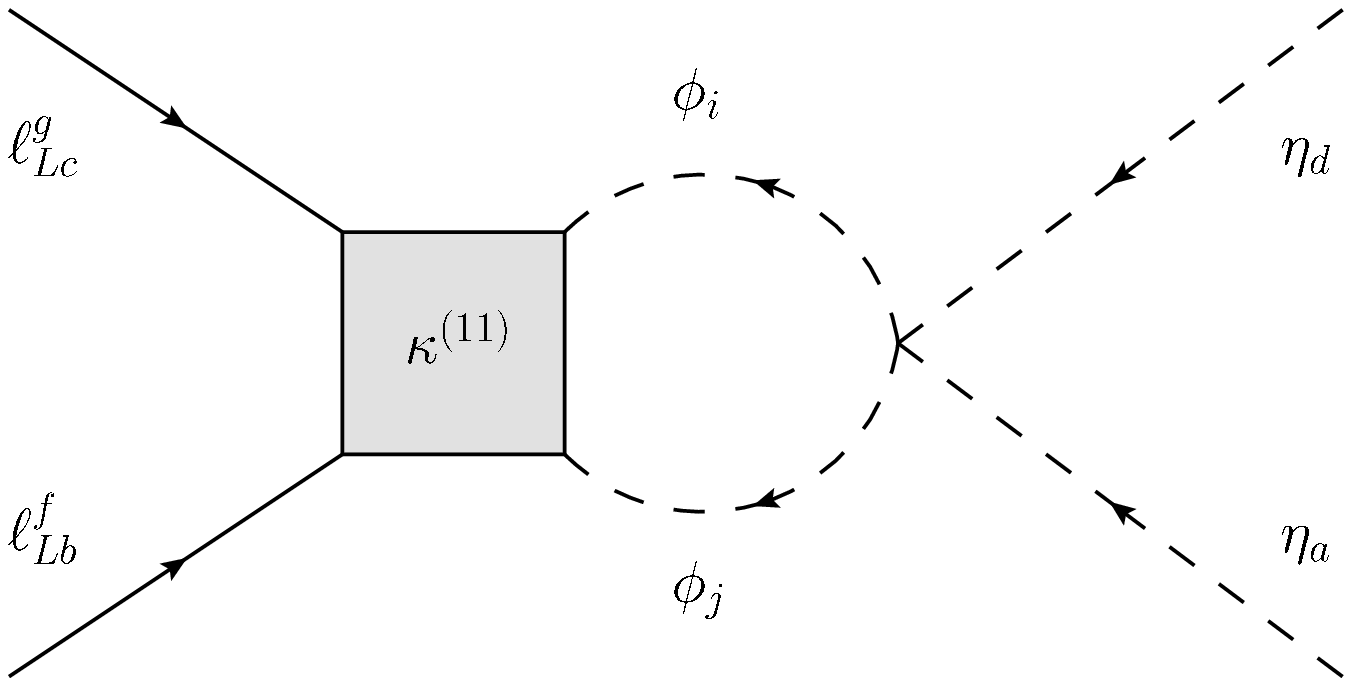}
\includegraphics[width=7cm]{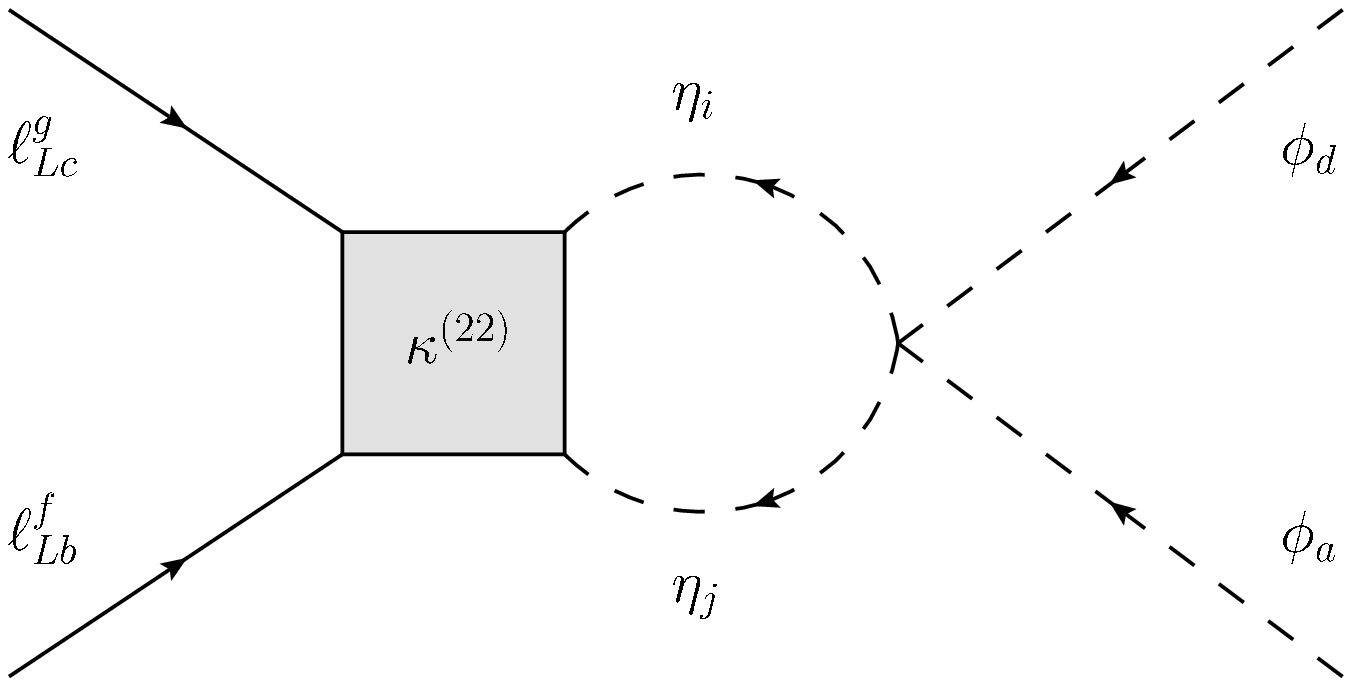}
\caption{\label{fig:eff_mix} The 1-loop mixing of the two Weinberg-like operators in the Ma-model.}
\end{figure}

To have a consistent way to describe the running of the neutrino masses, we hence make use of the so-called \emph{Weinberg operator}~\cite{Weinberg:1979sa},
\begin{equation}
\mathcal{L}_{\kappa}^{d=5}=\frac{1}{4}\kappa_{gf} \underbrace{\overline{\ell_{Lc}^{g\mathcal{C}}} \varepsilon_{cd}\phi_d\ell_{Lb}^f\varepsilon_{ba}\phi_a}_{\equiv \mathcal{O}_\kappa :\,\text{effective operator}} + \text{h.c.}\,,
\label{eq:smeff}
\end{equation}
where $\epsilon_{ij}$ is the two-dimensional antisymmetric tensor. Indeed, the 1-loop diagram in Fig.~\ref{fig:seesaws} contributes to the Weinberg operator, and it corresponds to the type {\bf T-3} contribution in the general classification in Ref.~\cite{Wein5}. With only the SM particle content, the Weinberg operator is the only $d=5$ operator possible, and it will give masses to neutrinos once the Higgs field obtains a VEV. However, in our case we have an additional scalar doublet $\eta$, which allows for more Weinberg-like operators:
\begin{equation}
\mathcal{L}_{\kappa}^{d=5}=\sum\limits_{r,s=1}^2 \mathcal{L}_{\kappa^{(rs)}}=\sum\limits_{r,s=1}^2\frac{1}{4}\kappa^{(rs)}_{gf} \underbrace{\overline{\ell_{Lc}^{g\mathcal{C}}} \epsilon_{cd}\phi_d^{(r)}\ell_{Lb}^f\epsilon_{ba}\phi_a^{(s)}}_{\equiv \mathcal{O}_\kappa^{(rs)}:\,\text{effective operators}} + \text{h.c.}\,,
\label{eq:sumeff}
\end{equation}
where $\phi^{(1)}=\Phi$ and $\phi^{(2)}=\eta$. These four potential operators are depicted in Fig.~\ref{fig:eff}. However, it is clear that the $Z_2$ symmetry of the Ma-model forbids two of these operators, $\mathcal{O}^{(12)}_\kappa$ and $\mathcal{O}^{(21)}_\kappa$, and thus the final remaining operators are $\mathcal{O}^{(11)}_\kappa$ and $\mathcal{O}^{(22)}_\kappa$. For convenience these operators will be written as $\kappa^{(11)}$ and $\kappa^{(22)}$, but one has to keep in mind that this is not purely correct, as the $\kappa$'s are actually just the coefficients of the $d=5$ operators, and not the operators themselves. Note that $\kappa^{(22)}$ will not directly contribute to the neutrino mass, as the inert scalar does not obtain a VEV. However, it will mix with the SM Weinberg operator at 1-loop level, cf.\ Fig.~\ref{fig:eff_mix}, making it necessary when attempting to impose a consistent matching at 1-loop level (cf.\ Ref.~\cite{Antusch:2001vn}). As a result, the RGEs of these two operators are coupled.

Now we know that our low-energy effective theory contains two effective operators (which are mixed at 1-loop level) contributing to the mass of the neutrinos. We can then renormalize the effective theory and evolve our set of parameters (Yukawa couplings, scalar couplings, scalar masses, effective operators) to the scale at which this theory breaks down, typically the mass of one of the heavy neutrinos. However, there is no reason why the three neutrino singlets would be degenerate in mass. Then, instead of integrating out all the heavy neutrinos at once to go directly from the full to the lowest-energy effective theory, it would seem more natural to include a certain hierarchy among the heavy neutrino masses and to integrate out the corresponding fields one by one. Consequently, we will obtain a set of three different effective theories characterized by their own respective scales, i.e., the three masses of the heavy neutrinos. These Majorana masses are unrelated to the EWSB scale and they will be considered to be of the order of several TeV. Later, the three effective theories will be denoted by EF1, EF2, and EF3, while the full theory will be denoted by FT.

\section{\label{sec:Ma-tching}Matching the theories}

Let us summarize what is needed to be done to get the global running of the parameters over the set of theories:
\begin{enumerate}

\item We start at a scale above the masses of all the particles, where the renormalizable theory (i.e., the FT) is valid.

\item The next step is to evolve this theory down to lower scales. As long as the energy scale under consideration lies above all heavy neutrino masses, this evolution is described by the RGEs of the FT.

\item When the energy scale considered goes below the mass $M_k$ of one of the neutrino singlets in the theory, we must change the theory to the new theory ET$k$ without the corresponding field $N_k$, thereby introducing non-renormalizable interactions via effective operators.

\item Both the changes in the existing parameters and in the coefficients of the new interactions are computed by matching the physics directly above and below the boundary between the two theories. This must be done at each mass scale boundary.

\end{enumerate}

To achieve the first three points, it is just necessary to renormalize each theory, taking into account the changes in the particle content and in the effective operators. We will end up with parameters whose running is described by piecewise continuous functions.  To avoid a mismatch at the junctions between theories, we need the fourth point: Indeed, as often in physics, the use of highly simplified models brings some unphysical behaviour. Here it is obvious that the heavy neutrinos do not decouple abruptly, but instead their influence weakens slowly until it becomes insignificant. The exact behaviour is complex, but we can fix the potential mismatch of the running between different theories by imposing so-called \emph{matching conditions}. This procedure has the advantage of being really straightforward, and it consists only in evaluating both theories at the matching scale, i.e., the energy scale at the boundary. This procedure is standard in the study of the running of the parameters when considering different effective theories (see, e.g., \cite{Antusch:2005gp,Antusch:2002rr,Bergstrom:2010id,Antusch:2001vn,Bergstrom:2010qb}). However, it is usually performed at tree-level, whereas we must consider it at 1-loop level to obtain a consistent answer, due to the 1-loop mixing between $\kappa^{(11)}$ and $\kappa^{(22)}$. The corresponding diagrams for the different effective theories are depicted in Tab.~\ref{tab:eff}. Power counting indicates that they do not all have the same degrees of divergence. Indeed, diagrams in ETs have typically higher degrees of UV divergence, as they contain fewer propagators. For example, in the case of $\kappa^{(11)}$, the second diagram in ET$3$ is logarithmically divergent, while the corresponding diagram in the FT is finite. This is not an obstacle, but we will have to regulate each diagram using dimensional regularization. Once that is done, the infinite part will be skipped, as it can be compensated by an adequate counterterm in the Lagrangian. Then, only the finite scale-dependent part will matter for the matching.  Note that, in our case, we have one more interesting simplification: Due to the 1-loop diagrams being proportional to some parameters $\lambda_i$ from the Higgs potential, which is not the case for the tree-level diagrams, we can separate the tree- and 1-loop level contributions and match them separately rather than matching their sums. The matching scheme for the tree-level diagrams is depicted in Tab.~\ref{tab:tree-match} for illustration.

\begin{table}[t]
\centering
\begin{tabular}{|c||c|c|}\hline
  & \raisebox{-2pt}{$\kappa^{(11)}$} & \raisebox{-2pt}{$\kappa^{(22)}$}\\ \hline \hline
\raisebox{1.76cm}{FT} & \includegraphics[width=0.24\textwidth]{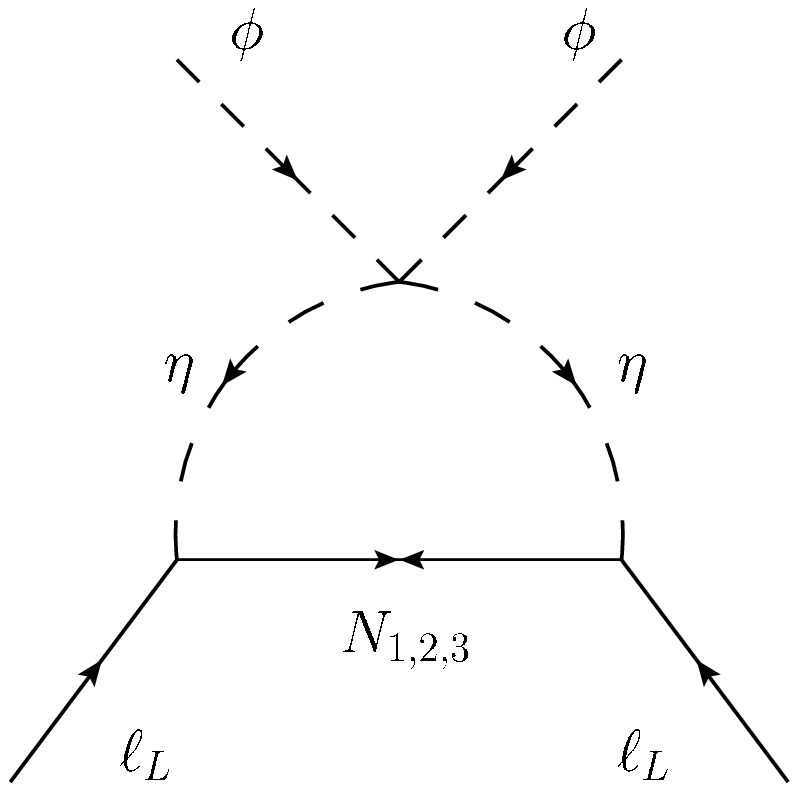} & \includegraphics[width=0.56\textwidth]{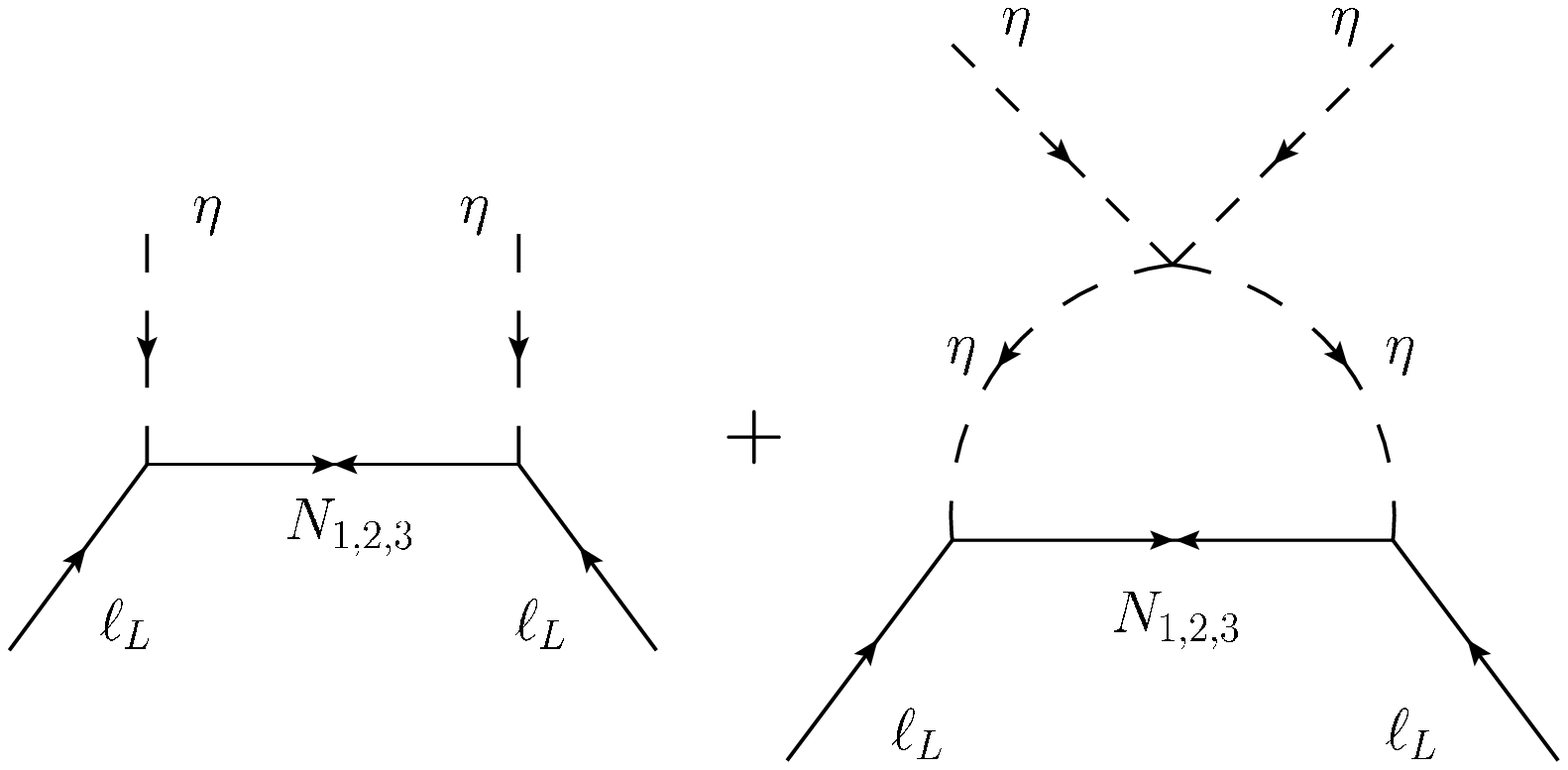}\\ \hline
\raisebox{2.64cm}{ET$3$} & \includegraphics[width=0.24\textwidth]{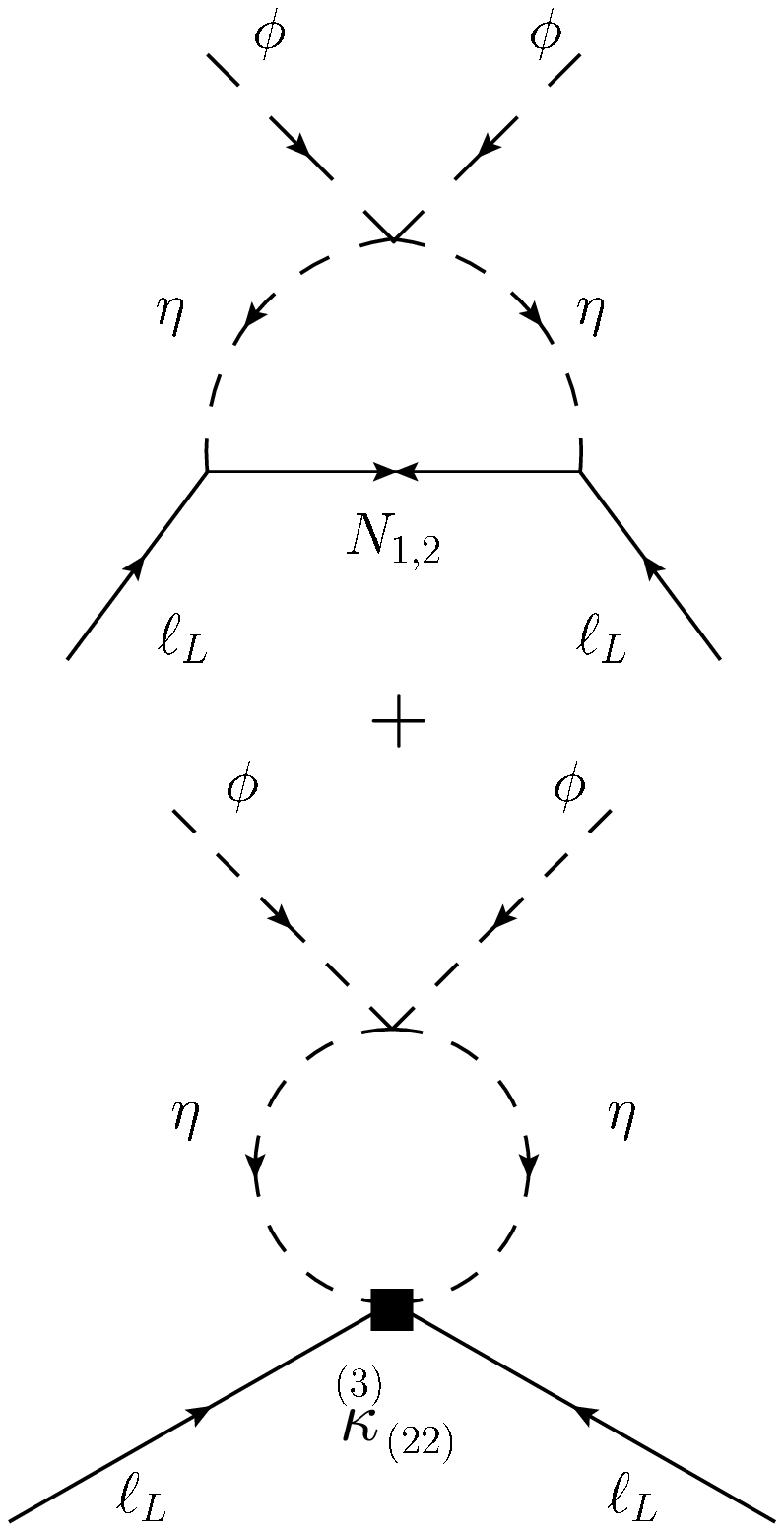} & \includegraphics[width=0.56\textwidth]{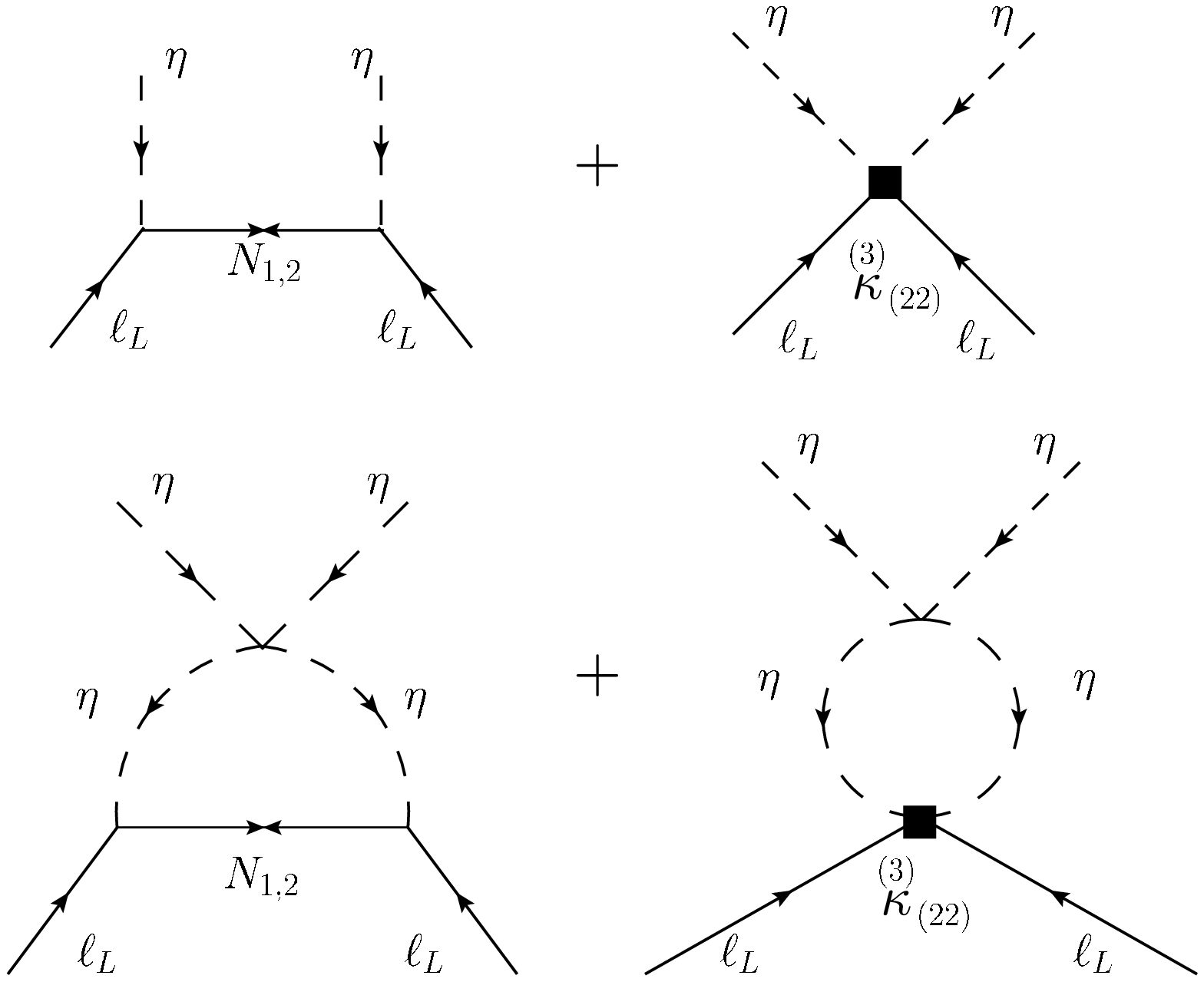} \\ \hline
\end{tabular}
\caption{\label{tab:eff}Overview of the diagrams to consider for the matching in the different theories (FT \& ET$3$). Effective operators are drawn in a space-saving version.}
\end{table}

\begin{table}[t]
\ContinuedFloat
\centering
\begin{tabular}{|c||c|c|}\hline
  & \raisebox{-2pt}{$\kappa^{(11)}$} & \raisebox{-2pt}{$\kappa^{(22)}$}\\ \hline \hline
\raisebox{2.64cm}{ET$2$} & \includegraphics[width=0.24\textwidth]{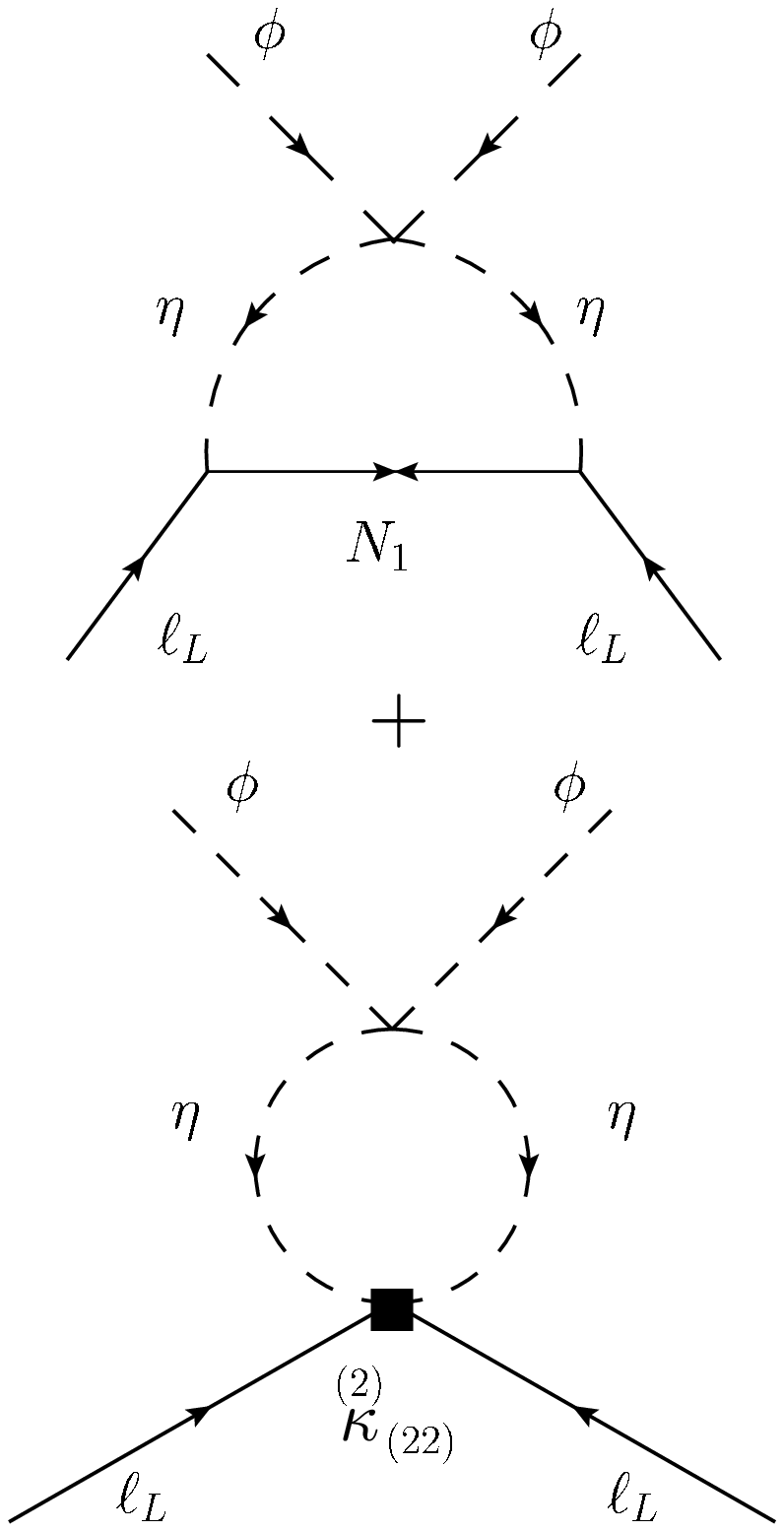} & \includegraphics[width=0.56\textwidth]{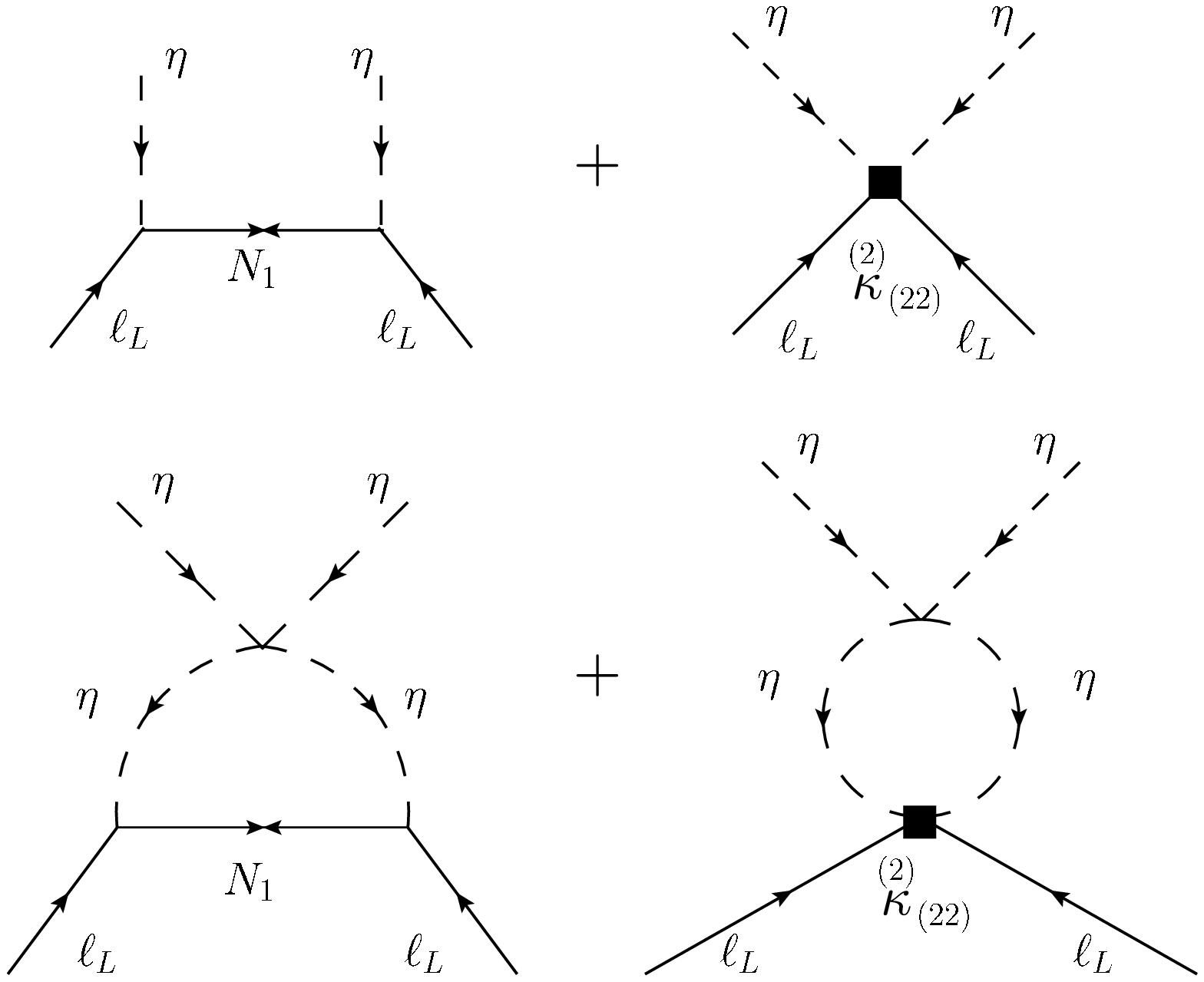}\\ \hline
\raisebox{1.76cm}{ET$1$} & \includegraphics[width=0.28\textwidth]{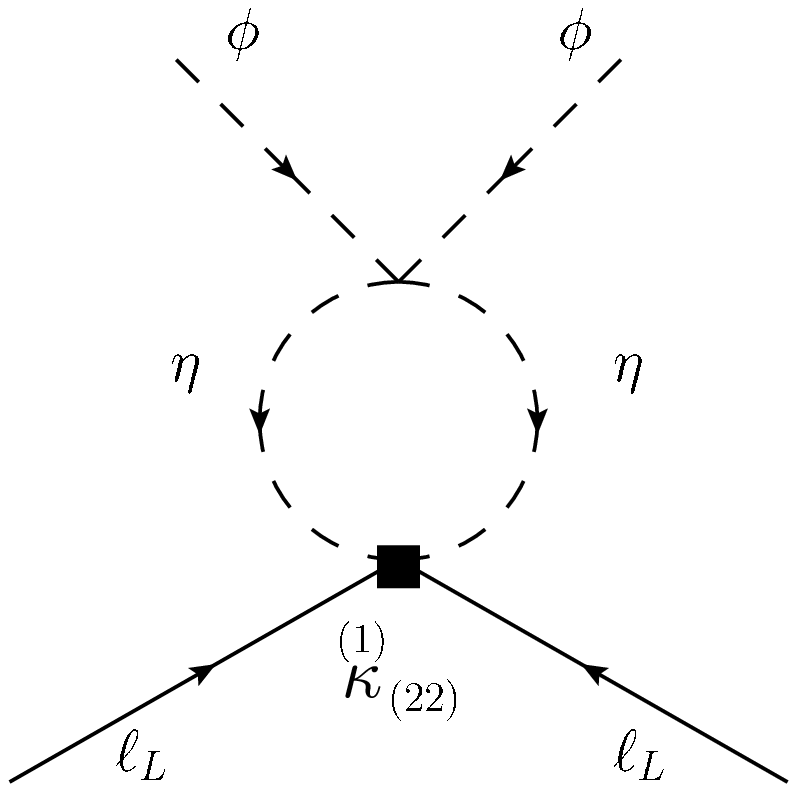} & \includegraphics[width=0.52\textwidth]{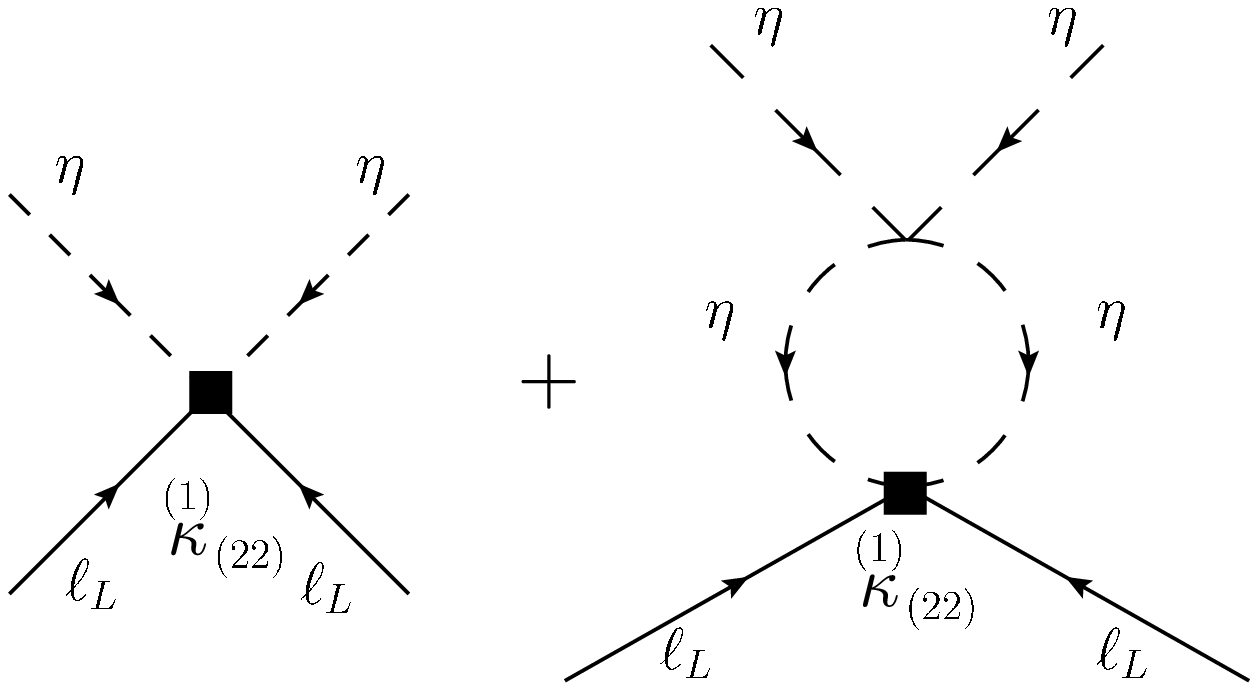} \\ \hline
\end{tabular}
\caption{\label{tab:eff_2}(\emph{continued}) Overview of the diagrams to consider for the matching in the different theories (ET$2$ \& ET$1$). Effective operators are drawn in a space-saving version.}
\end{table}

\begin{table}[!h]
\centering
\begin{align}
&\raisebox{22pt}{FT}
\includegraphics[width=0.3\textwidth]{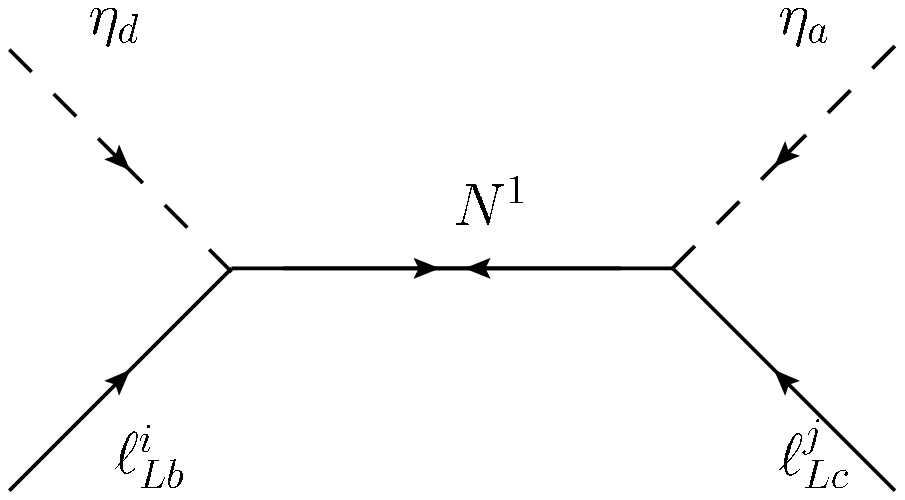}
\raisebox{22pt}{+}\includegraphics[width=0.3\textwidth]{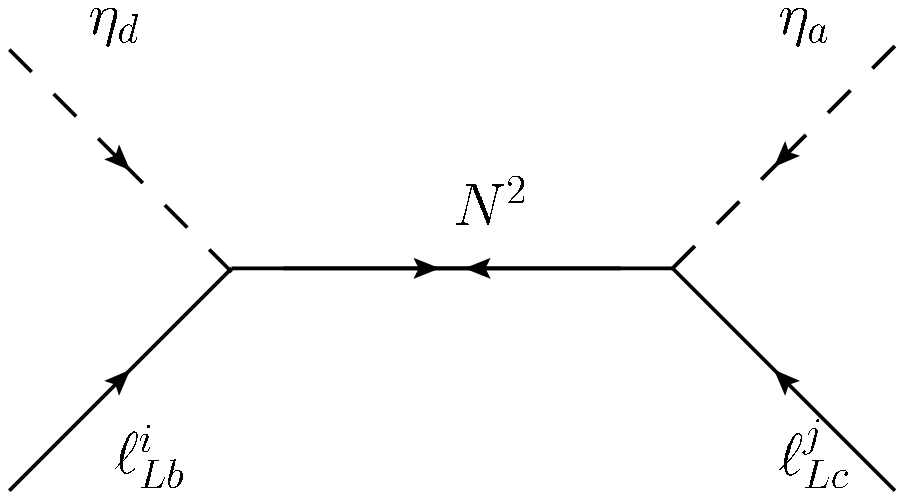}
\raisebox{22pt}{+}\underbrace{\includegraphics[width=0.3\textwidth]{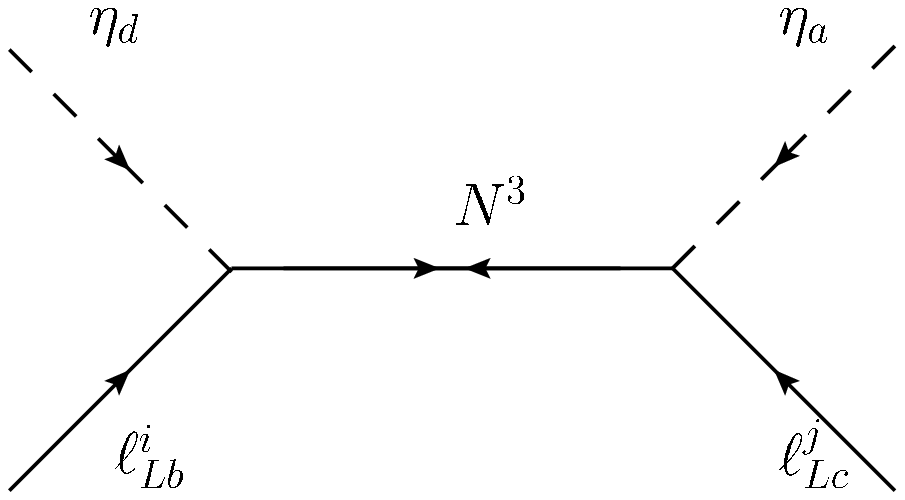}}_{\qquad\quad\downarrow\quad \text{MC3}}\nonumber\\
&\raisebox{22pt}{ET3}
\quad\includegraphics[width=0.3\textwidth]{NewScheme1}
\raisebox{22pt}{+}\underbrace{\includegraphics[width=0.3\textwidth]{NewScheme2} \raisebox{22pt}{+} \includegraphics[width=0.2\textwidth]{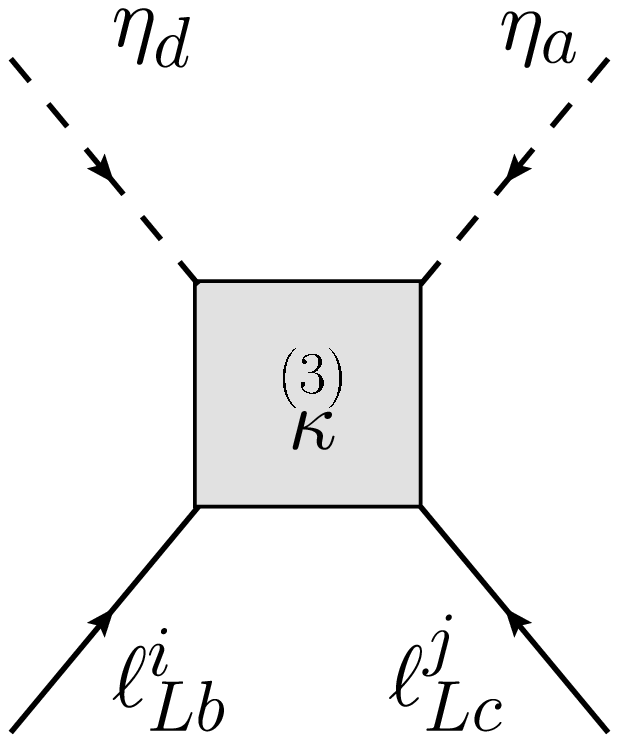}}_{\qquad\quad\downarrow\quad \text{MC2}}\nonumber\\
&\raisebox{22pt}{ET2}
\quad\qquad\qquad\qquad\underbrace{\includegraphics[width=0.3\textwidth]{NewScheme1} \raisebox{22pt}{+} \includegraphics[width=0.2\textwidth]{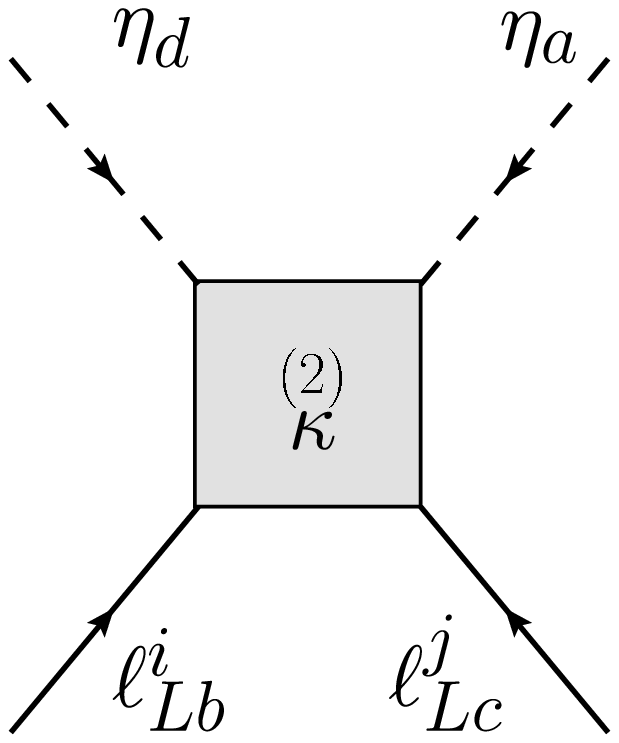}}_{\qquad\quad\downarrow\quad \text{MC1}}\nonumber\\
&\raisebox{22pt}{ET1}
\quad\qquad\qquad\qquad\qquad\qquad\qquad \includegraphics[width=0.2\textwidth]{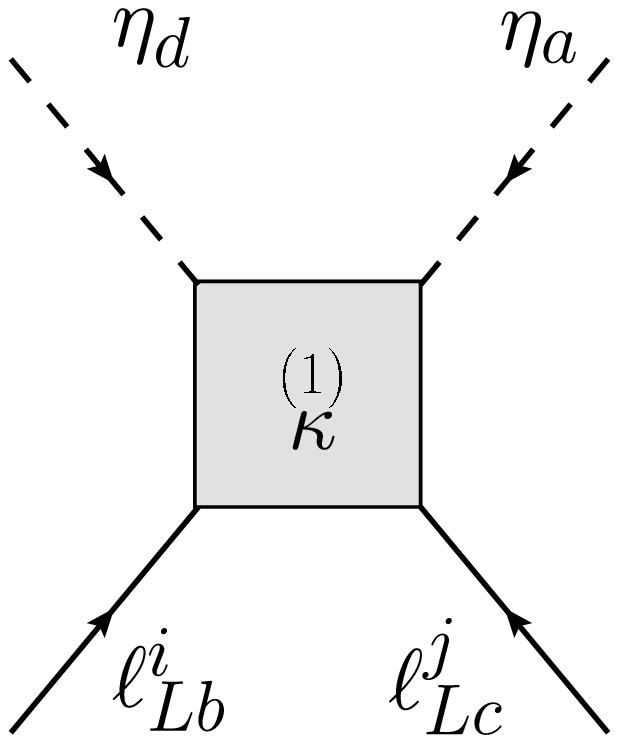}\nonumber
\end{align}
\caption{\label{tab:tree-match} The matching of the tree-level diagrams. None of these operators can directly lead to a light neutrino mass, but they will contribute at 1-loop level. The abbreviation ``MC'' stands for matching condition.}
\end{table}

We have performed the calculations of all these diagrams, and of all other diagrams relevant for the running. This includes several running quantities such as gauge couplings or parameters from the Higgs potential, above the energy scale of EWSB, where only the scalar doublets and the neutrino singlets have mass. Note that, at the lowest energies, this actually introduces a slight inconsistency: EWSB would set in, all particles would become massive, and in particular we might have to integrate out some of the inert scalars. However, that would happen at a scale very close to $M_Z$, and unless the threshold effects are extremely strong the resulting effect would be negligible. We will leave the detailed investigation of this parameter region for future studies, as the main results presented in this paper are the RGEs and the discussion of one very illustrative case.

We now impose the matching conditions. This is a systematic procedure which is done order by order in the loop-expansion. When two theories are compared at a given loop order, the lower order results have to be included in the matching. In our special case, however, they happen to decouple. Furthermore, despite of having a stack of three effective theories, going from one to another is each time a similar process, namely the decoupling of exactly one heavy neutrino (unless two or more of them were degenerate in mass). Therefore, the matching conditions will be analogous and we can write down a general expression, valid whichever the ET relevant for the energy scale is considered. For simplicity, we adopt the convenient notations from Ref.~\cite{Antusch:2005gp}:
\begin{itemize}

\item $\kappa^{(rr)}\longrightarrow \overset{\scriptscriptstyle (n)}{\kappa}_{(rr)}$ is the effective operator in the $n$-th effective theory, i.e., the operator making up for the low-energy influence of the heavy neutrinos $N_k$ (with $k=n,\ldots,3$).

\item $M_N\longrightarrow \overset{\scriptscriptstyle (n)}{M}_N$ is the mass matrix of the remaining heavy neutrino singlets in ET$n$. In the FT this matrix is of dimension $3\times 3$, but in ET$n$ the states $N_k$ with $k\geq n$ have been decoupled, and their masses are not defined anymore. As a result the heavy neutrino Majorana mass matrix in ET$n$ is an $(n-1)\times (n-1)$ matrix. We denote by $M_n$ the largest eigenvalue of $\smas{n+1}$.

\item $h_\nu\longrightarrow \overset{\scriptscriptstyle (n)}{h}_\nu$ is the neutrino Yukawa matrix taking in account that we have integrated out $(4-n)$ generations of neutrino singlets. It implies that it is a non-square matrix coupling of $(n-1)$ neutrino singlets to 3 lepton doublets. Taking rows corresponding to neutrino singlets and columns corresponding to lepton doublets, the coupling technically looks like an $(n-1)\times 3$ submatrix of the ``complete'' Yukawa matrix:
\vspace{5pt}
\[
h_\nu\equiv
\begin{array}{clllll}\label{tabernacle}
  \ldelim({8}{0.3cm} &(h_\nu)_{1,1} 		& (h_\nu)_{1,2} 		& (h_\nu)_{1,3} 	&\rdelim){8}{0.3cm} &\rdelim\}{4}{2pt}[$\,\rightarrow\overset{\scriptscriptstyle{(n)}}{h}_\nu$]\\[3pt]
  &\qquad\vdots 		& \qquad\ddots		& \qquad\vdots		& &\\[3pt]
  &(h_\nu)_{n-1,1}  	& (h_\nu)_{n-1,2} 	& (h_\nu)_{n-1,3}	& &\\[3pt]
  &\qquad 0				& \qquad\cdots 		& \qquad 0			& &\\[3pt]
  &\qquad\vdots			& \qquad\ddots		& \qquad\vdots		& &\\[5pt]
  & \qquad 0			& \qquad\cdots		& \qquad 0			& & 
\end{array}
\]
\vspace{0.3cm}
\end{itemize}
Note that with these notations, we have:
\begin{equation}
\overset{\scriptscriptstyle (1)}{\kappa}_{(rr)}=\kappa^{(rr)}, \ \ \ \overset{\hspace{-0.5cm}\scriptscriptstyle (n>3)}{\kappa_{(rr)}}=0,\ \ \ \suph{4}=h_\nu,\ \ \ \suph{1}=0.
\end{equation}
In practice we merely match the values of the diagrams, which change when going from one theory to another. Considering the scheme from above and performing the calculation, one obtains the following matching conditions (with $n=1,2,3$):
\begin{enumerate}
\item For $\kappa^{(11)}$:
\begin{equation}
\text{MC}n\,(\mu=M_n) \longrightarrow \,\keffi{n}\ =\ \keffi{n+1}+\frac{f(M_n,m_2)}{g(M_n,m_2)}(\supht{n+1})_{in}M^{-1}_n(\suph{n+1})_{nj}\,.
\end{equation}
\item For $\kappa^{(22)}$:
\begin{equation}
\text{MC}n\,(\mu=M_n) \longrightarrow \,\keffid{n}\ =\ \keffid{n+1}+\left[2+\frac{f(M_n,m_2)}{g(M_n,m_2)}\right] (\suph{n+1})_{in}M^{-1}_n(\supht{n+1})_{nj}\,.
\end{equation}
\end{enumerate}
Here, we have used the loop functions and $f(x,y)\equiv\frac{x^4}{(x^2-y^2)^2}\ln\left(\frac{x^2}{y^2}\right)+\frac{x^2}{x^2-y^2}$ and $g(x,y)\equiv\ln\left(\frac{y^2}{x^2}\right)$, which arise from the computations of the corresponding diagrams~\cite{ThesisRomain}. In general, $\mu$ denotes the renormalization scale.

There is one more subtlety involved: The matching conditions, as presented here, are valid only in the basis where the heavy neutrino Majorana mass matrix is diagonal, which is the setup we imposed at the input scale. However, the running does not care for our choice of basis and it will introduce some off-diagonal coefficients in our initially diagonal matrices~\cite{Antusch:2002rr}. Therefore, each time when we integrate out one heavy neutrino at the corresponding threshold, the Majorana mass matrix will have to be diagonalized first, so that we can use the derived expressions. This procedure imposes a redefinition of the Yukawa matrix that can be written as follows:
\begin{equation}
\smas{n}\quad \to \quad V\overset{\scriptscriptstyle{(n)}}{M_{\rm diag}}V^T,\ \ \ \suph{n}\quad\to\quad V^T\suph{n}.
\end{equation}
Here, $V$ must be a unitary matrix since $\smas{n}$ is Hermitian. We have to perform this ``re-diagonalization'' at each matching scale.

Finally, we can write down the most general expression for the light neutrino mass matrix, arising from all contributions (only at 1-loop in this case) to the operator $\kappa^{(11)}$:
\begin{equation}
\mathcal{M}_\nu(\mu)=v^2 \left(\frac{\lambda_5}{16\pi^2}\right)\left[g(\mu,m_2) \overset{\scriptscriptstyle (n)}{\kappa}_{(22)} +\supht{n}\ \overset{\hspace{-0.3cm}\scriptscriptstyle (n)}{M_N^{-1}} f(M_n,m_2)\suph{n}\right].
\label{eq:numass-gen}
\end{equation}
Note that the structure of this matrix is actually very similar to the one in an ordinary seesaw type~I framework (or in the corresponding effective theories, respectively), but corrected by some loop functions. To determine the leptonic mixing, it is enough to diagonalize this light neutrino mass matrix with a unitary matrix $U_\nu(\mu)$, as well as the charged lepton mass matrix with a unitary matrix $U_e(\mu)$. Note that these matrices, naturally, depend on the energy scale $\mu$ under consideration, which will in the end translate into the running of the neutrino mixing parameters. Hence, the Pontecorvo-Maki-Nagakawa-Sakata (PMNS) leptonic mixing matrix is directly given by:
\begin{equation}
U_{\rm PMNS}(\mu)= U_e^\dagger(\mu)U_\nu(\mu).
\end{equation}
We adopt for $U\equiv U_{\rm PMNS}$ the following parametrization:
\begin{equation}
U=
\begin{pmatrix}
c_{12}c_{13}					&s_{12}c_{13}		&s_{13}e^{-i\delta}\\
-s_{12}c_{23}-c_{12}s_{23}s_{13}e^{i\delta} &c_{12}c_{23}-s_{12}s_{23}s_{13}e^{i\delta}&s_{23}c_{13} \\ s_{12}c_{23}-c_{12}s_{23}s_{13}e^{i\delta}  &-c_{12}c_{23}-s_{12}s_{23}s_{13}e^{i\delta}&c_{23}s_{13}
\end{pmatrix}\begin{pmatrix}e^{i\rho} &0 &0\\0&e^{i\sigma}&0\\0&0&1\end{pmatrix},
\end{equation}
where $s_{ij} = \sin \theta_{ij}$ and $c_{ij} = \cos \theta_{ij}$. The mixing parameters can then be extracted from the PMNS-matrix by~\cite{Ray:2010rz}:
\begin{subequations}
\begin{align}
\theta_{13} &=\arcsin (|U_{13}|)\,,\label{eq:mixpar_a}\\
\theta_{12} &=\left\{
  \begin{array}{ l l l}
    &\arctan\left( \dfrac{|U_{12}|}{|U_{11}|}\right)  & \quad \text{if $U_{11}\neq 0$}\\
    & \frac{\pi}{2} & \quad \text{else}\\
  \end{array} \right. ,  \label{eq:mixpar_b}\\
\theta_{23} &=\left\{
  \begin{array}{ l l l}
    &\arctan\left( \dfrac{|U_{23}|}{|U_{33}|}\right)  & \quad \text{if $U_{33}\neq 0$}\\
    & \frac{\pi}{2} & \quad \text{else}\\
  \end{array} \right. ,  \label{eq:mixpar_c}\\
\delta \quad &=-\text{arg}\left[ \left( \dfrac{U_{11}^\ast U_{13}U_{31}U_{33}^\ast}{c_{12}c_{13}^2c_{23}s_{13}}+c_{12}c_{23}s_{13} \right) / \left( s_{12}s_{23} \right) \right]. \label{eq:mixpar_d}
\end{align}
\end{subequations}
Now that we have discussed the principle ideas, we can move on to our results, by first presenting the analytical forms of the RGEs and then investigating their numerical solutions in a simplified example scenario.

\section{\label{sec:RGEs}One-loop Renormalization Group Equations}

The main result of this work is the set of 1-loop RGEs for Ma's scotogenic neutrino mass model, to be presented in this section. The presentation of this full set of RGEs (with the only exception of the effective theory with the scalars being integrated out) is, to our knowledge, a completely new result. However, of course certain limiting cases have already been discussed in the literature.

Renormalization group equations have been extensively studied, e.g., in the contexts of the SM and the MSSM. The references to these are numerous and sometimes messy, so we will refer to the reader only to the very nice Refs.~\cite{KerstenPhD,Chankowski:2001mx,Pirogov:1998tj} for the presentation of an exhaustive set of RGEs. However, in the theoretical context of the model investigated here, these equations are not adequate, and the best way to find consistent results is to consider the derivation of the RGEs in a general quantum field theory, as performed in Refs.~\cite{Machacek1,Machacek2,Machacek3}, or in a general THDM~\cite{Cheng:1973nv}.

More specifically, type~I seesaw models have been discussed in Refs.~\cite{Antusch:2005gp,Antusch:2002rr,kerstendipl}. Tree-level extensions of the seesaw mechanism have been discussed in Refs.~\cite{Schmidt:2007nq,Bergstrom:2010id,Bergstrom:2010qb,Ray:2010rz}, and the case of Two- (or Multi-) Higgs Doublet Models has been presented in Refs.~\cite{Antusch:2005gp,Ibarra:2011gn}. Stability of texture zeros has been investigated in Refs.~\cite{KerstenPhD,Hagedorn:2004ba}. Studies of the Weinberg operator (or Weinberg-like operators) have been performed, e.g., in the context of the SM~\cite{Antusch:2001ck} or in a general THDM~\cite{Antusch:2001vn}. Finally, extremely useful software packages have been developed in connection with Ref.~\cite{Antusch:2005gp}: {\sf REAP} (``Renormalization Group Evolution of Angles and Phases'')~\cite{REAP} and {\sf MPT} (``Mixing Parameter Tools'')~\cite{MPT}, to be found online. Whenever applicable, we have checked the consistency of our results with these references, and we have always been able to reach agreement.

Hereafter we present the set of RGEs corresponding of the Ma-model, where throughout the paper we will use the abbreviation
\begin{equation}
\mathcal{D}\equiv 16\pi^2 \mu \frac{\mathrm{d}}{\mathrm{d}\mu}, 
\end{equation}
with $\mu$ being the renormalization scale. Then, the following parameters will be involved in the running, and their running (in the FT) is determined most easily in the order indicated here:
\begin{enumerate}

\item The gauge couplings $g_i$, where $i=1,2,3$.

\item The SM-like Yukawa coupling matrices $Y_x$, with $x=u,d,e$, for the up-quarks, down-quarks, and charged leptons, as well as the neutrino Yukawa coupling matrix in the Ma-model, $h_\nu$.

\item The heavy neutrino Majorana mass matrix $M_N$ and the five real scalar couplings $\lambda_i$, with $i=1,2,3,4,5$, which appear in the Higgs potential.

\item The mass parameters $m_1$ and $m_2$ of the SM-like Higgs and of the inert scalar, respectively.

\end{enumerate}

In addition to these quantities, we also have the effective operators $\overset{\scriptscriptstyle (n)}{\kappa}_{(11)}$ and $\overset{\scriptscriptstyle (n)}{\kappa}_{(22)}$ running in the different effective theories ET$n$, whose RGEs are coupled, cf.\ Sec.~\ref{sec:Ma-EFT}. All these RGEs have been calculated using dimensional regularization, while adopting the minimal subtraction (MS) scheme. We will now present the 1-loop RGEs and discuss the different sets of equations one by one. While in this paper we present the resulting equations, more details on their derivation and on the actual calculation can be found in the corresponding thesis, Ref.~\cite{ThesisRomain}.

\subsection{\label{sec:RGE_gauge}The gauge couplings}

\begin{figure}[t]
\centering
\includegraphics[width=7.9cm]{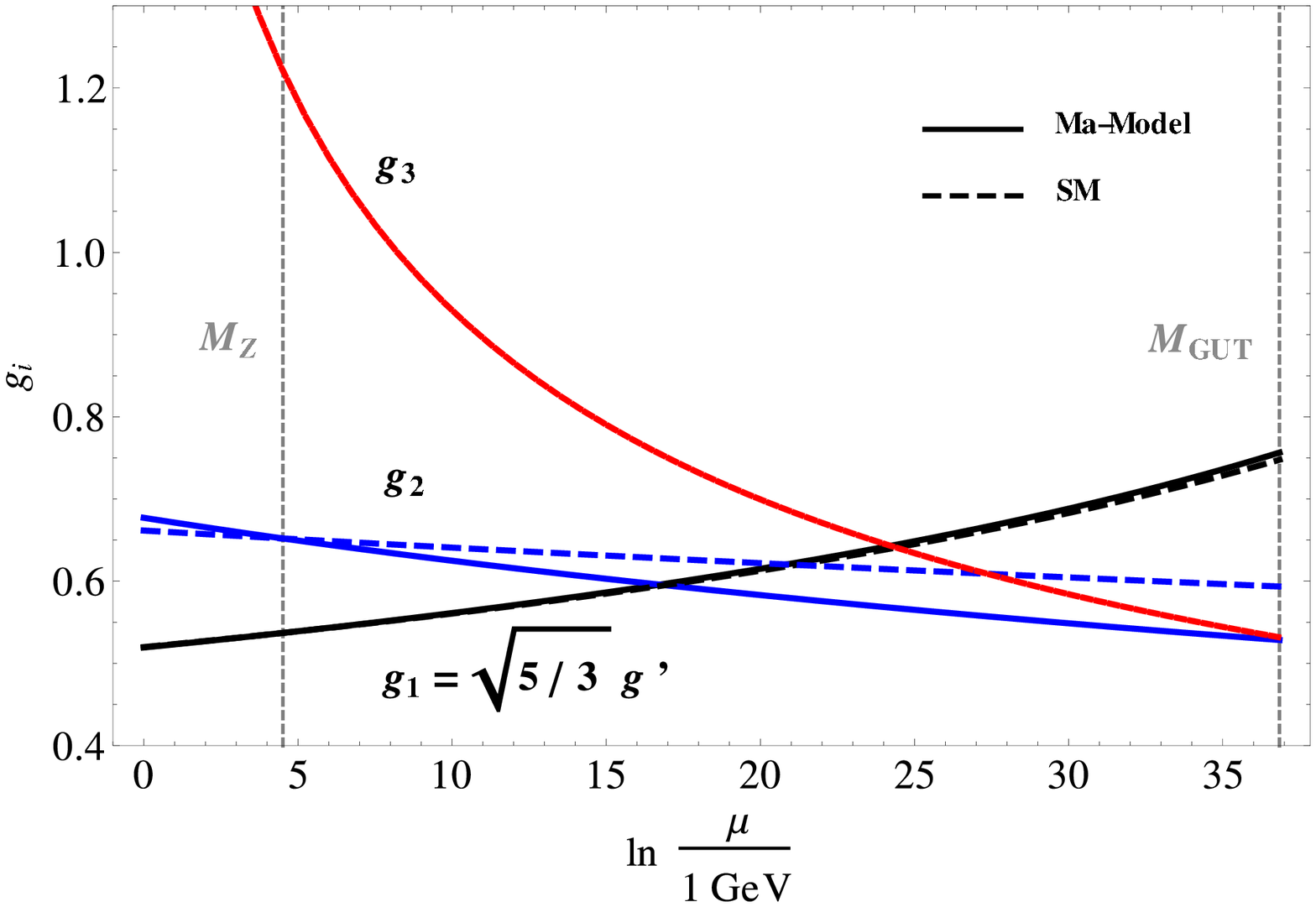}
\includegraphics[width=7.9cm]{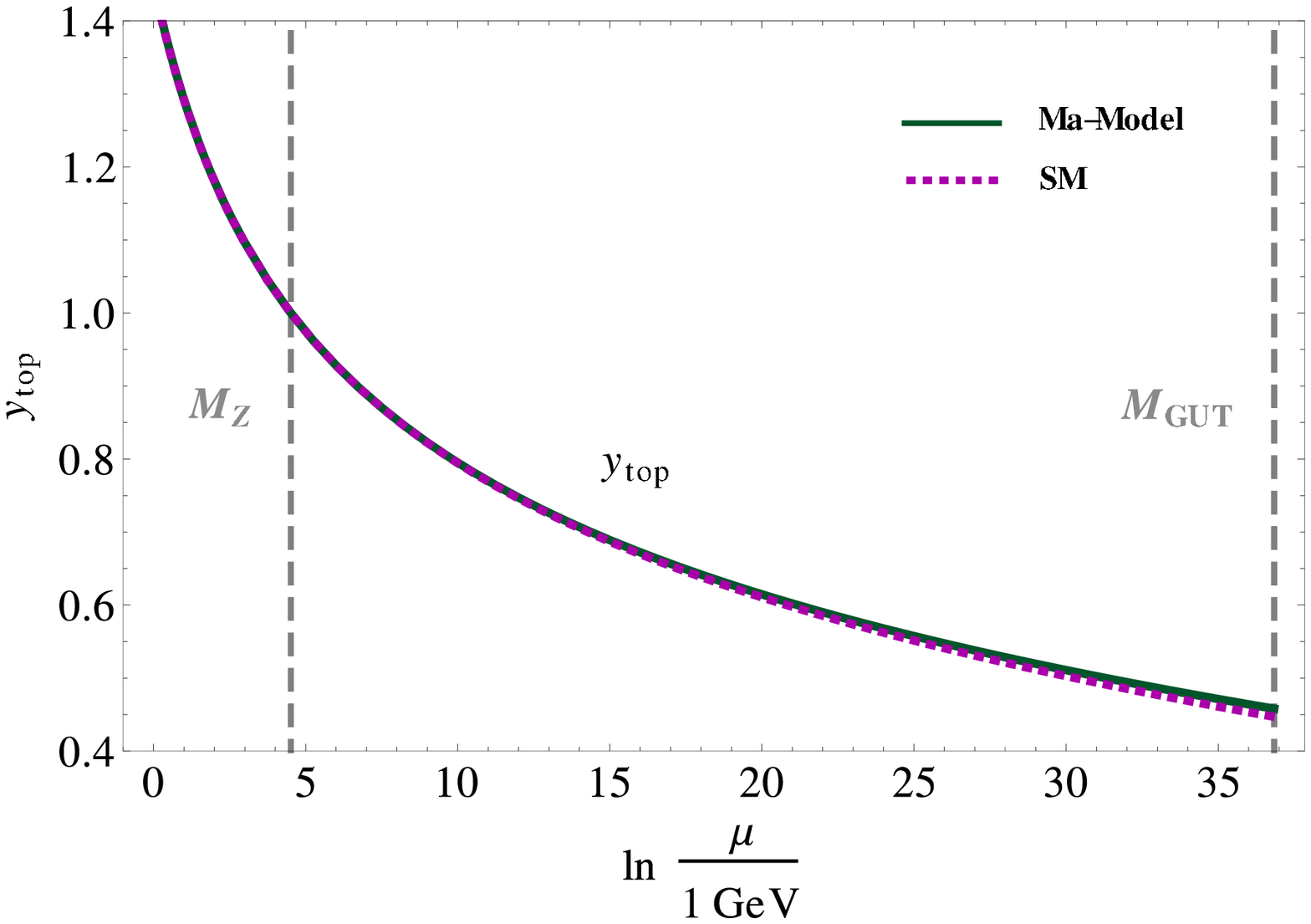}
\caption{\label{fig:SM-coup} The runnings of the gauge couplings and of the top Yukawa coupling in the SM and in the Ma-model.}
\end{figure}

The RGEs for the gauge couplings are given by:
\begin{subequations}
\begin{align}
\mathcal{D}g_1 &= 7 g_1^3\label{eq:gauge_RGE_a},\\
\mathcal{D}g_2 &= -3 g_2^3\label{eq:gauge_RGE_b},\\
\mathcal{D}g_3 &= -7 g_3^3\label{eq:gauge_RGE_c}.
\end{align}
\end{subequations}
Compared to the SM, these equations are only slightly altered because they do not depend on the Yukawa and scalar couplings but only on the number of fermion generations and the number of scalar doublets in the theory~\cite{Grzadkowski198764}. Note that the differential equations~\eqref{eq:gauge_RGE_a} to~\eqref{eq:gauge_RGE_c} can all be easily solved analytically, leading to
\begin{equation}
 g_i (\mu) = \frac{g_i(\mu_0)}{\sqrt{1-\frac{b_i}{8 \pi^2} g_i^2(\mu_0) \ln \left( \frac{\mu}{\mu_0} \right) }},
 \label{eq:gauge_RGE_sol}
\end{equation}
where $(b_1,b_2,b_3)=(7,-3,-7)$ and $\mu_0$ denotes the reference scale. The running of the gauge couplings is depicted on the left panel of Fig.~\ref{fig:SM-coup}. In particular the running of $g_2$ shows a considerable difference in the Ma-model compared to the SM, due to the presence of an additional scalar doublet. Note that we have chosen the input values for the gauge couplings such that the measured values are correctly reproduced at the $Z$-pole, i.e., at the scale $\mu = M_Z$.

\subsection{\label{sec:RGE_Yukawa}The Yukawa couplings}

The RGEs for the Yukawa coupling matrices are given by:
\begin{subequations}
\begin{eqnarray}
\mathcal{D}Y_u &=& Y_u\left\lbrace \frac{3}{2}Y_u^\dagger Y_u -\frac{3}{2}Y_d^\dagger Y_d + T -\frac{17}{12} g_1^2-\frac{9}{4} g_2^2-8g_3^2\right\rbrace\label{eq:Yuk_RGE_a},\\[10pt]
\mathcal{D}Y_d &=& Y_d\left\lbrace \frac{3}{2}Y_d^\dagger Y_d -\frac{3}{2}Y_u^\dagger Y_u + T -\frac{5}{12} g_1^2-\frac{9}{4} g_2^2-8g_3^2\right\rbrace\label{eq:Yuk_RGE_b},\\[10pt]
\mathcal{D}Y_e &=& Y_e\left\lbrace \frac{3}{2}Y_e^\dagger Y_e +\frac{1}{2}h_\nu^\dagger h_\nu + T -\frac{15}{4} g_1^2-\frac{9}{4} g_2^2\right\rbrace\label{eq:Yuk_RGE_c},\\[10pt]
\mathcal{D}h_\nu &=& h_\nu\left\lbrace \frac{3}{2}h_\nu^\dagger h_\nu +\frac{1}{2}Y_e^\dagger Y_e + T_\nu -\frac{3}{4} g_1^2-\frac{9}{4} g_2^2\right\rbrace\label{eq:Yuk_RGE_d},
\end{eqnarray}
\end{subequations}
where $T\equiv \mathrm{Tr}\left(Y_e^\dagger Y_e +3Y_u^\dagger Y_u +3Y_d^\dagger Y_d\right)$ and $T_\nu\equiv \mathrm{Tr}\left(h_\nu^\dagger h_\nu\right)$. Note that, compared to the case of the SM supplemented by a Dirac neutrino Yukawa coupling matrix $Y_\nu$, the neutrino Yukawa coupling matrix $h_\nu$ in the Ma-model only appears in Eqs.~\eqref{eq:Yuk_RGE_c} and~\eqref{eq:Yuk_RGE_d}. However, numerically Eqs.~\eqref{eq:Yuk_RGE_a} and~\eqref{eq:Yuk_RGE_b} will not really differ from their equivalents in the SM with Dirac neutrino masses, due to the extremely small entries in the Yukawa matrix $Y_\nu$. The running of the top Yukawa coupling, in the approximation of this being the only non-zero entry of $Y_u$, is depicted on the right panel of Fig.~\ref{fig:SM-coup}. As to be expected, the top-running is dominated by the diagrams involving strongly interacting particles, and hence it is not really different in the Ma-model compared to the SM. Note that the set of equations~\eqref{eq:Yuk_RGE_a} to~\eqref{eq:Yuk_RGE_d} can be solved independently of all other parameters but the gauge couplings. Hence, by inserting the solutions of the equations from Sec.~\ref{sec:RGE_gauge}, the complete evolution of the Yukawa couplings can be studied.

\subsection{\label{sec:RGE_RH}The right-handed neutrino mass matrix}

The RGE required for the running of the mass matrix of the neutrino singlets can be found, e.g., in Refs.~\cite{Antusch:2002rr,Casas:1999ac}:
\begin{equation}
\mathcal{D}M_N = \left\lbrace M_N\left(h_\nu h_\nu^\dagger \right)^T+\left(h_\nu h_\nu^\dagger \right)M_N\right\rbrace.
\label{eq:N-evol}
\end{equation}
This equation is quasi identical to its seesaw type~I equivalent, cf.\ Ref.~\cite{kerstendipl}. Note that the only ``external'' input in this RGE is the Yukawa coupling matrix of the neutrinos, $h_\nu$. Hence, once the solution to Eq.~\eqref{eq:Yuk_RGE_d} is known, we can immediately obtain the numerical evolution of $M_N$.

\subsection{\label{sec:RGE_scalars}The scalar sector}

The RGEs of the scalar couplings have been calculated for the SM and its extended version including heavy neutrinos~\cite{Antusch:2005gp,KerstenPhD,Pirogov:1998tj}. In the context of a theory with more than two scalar doublets there are less studied, but some references give very general results~\cite{Cheng:1973nv}. However, in our specific case involving an additional scalar doublet as well as three neutrino singlets and an exact $Z_2$ symmetry, these results are not adequate. Thus, we had to adapt the RGEs to our case, leading to:

\begin{subequations}
\begin{eqnarray}
\mathcal{D}\lambda_1 &=& 12\lambda_1^2 +4\lambda_3^2 +4\lambda_3\lambda_4 +2\lambda_4^2 +2\lambda_5^2 + \frac{3}{4}\left( g_1^4+3g_2^4+2g_1^2 g_2^2\right) \label{eq:scal_run_a} \\
&& \qquad -3\lambda_1\left[ g_1^2+3g_2^2 -\frac{4}{3} T \right] -4 T_4, \nonumber\\
\mathcal{D}\lambda_2 &=& 12\lambda_2^2 +4\lambda_3^2 +4\lambda_3\lambda_4 +2\lambda_4^2 +2\lambda_5^2 + \frac{3}{4}\left( g_1^4+3g_2^4+2g_1^2 g_2^2\right) \nonumber \\
&& \qquad -3\lambda_2\left[ g_1^2+3g_2^2 -\frac{4}{3} T_\nu  \right]  -4 T_{4\nu}, \label{eq:scal_run_b}\\
\mathcal{D}\lambda_3 &=& \left( \lambda_1 + \lambda_2\right) \left( 6\lambda_3 +2\lambda_4\right)+4\lambda_3^2 +2\lambda_4^2 +2\lambda_5^2 + \frac{3}{4}\left( g_1^4+3g_2^4 - 2g_1^2 g_2^2\right) \nonumber \\
&& \qquad -3\lambda_3\left[ g_1^2+3g_2^2 -\frac{4}{3}\left(T_\nu + T\right)\right] -4 T_{\nu e}, \label{eq:scal_run_c}\\
\mathcal{D}\lambda_4 &=& 2\left( \lambda_1 + \lambda_2\right)\lambda_4 +8\lambda_3 \lambda_4 +4\lambda_4^2 +8\lambda_5^2 + 3 g_1^2 g_2^2 \nonumber \\
&& \qquad -3\lambda_4\left[ g_1^2+3g_2^2-\frac{2}{3}\left(T_\nu + T\right)\right] +4T_{\nu e},\label{eq:scal_run_d}\\
\mathcal{D}\lambda_5 &=& 2\left( \lambda_1 + \lambda_2\right)\lambda_5 +8\lambda_3 \lambda_5  +12\lambda_4\lambda_5 -3\lambda_5\left[ g_1^2+3g_2^2-\frac{2}{3}\left(T_\nu + T\right)\right],
\label{eq:scal_run_e}
\end{eqnarray}
\end{subequations}

\noindent
where $T_4\equiv \mathrm{Tr}\left\lbrace Y_e^\dagger Y_eY_e^\dagger Y_e +3Y_u^\dagger Y_uY_u^\dagger Y_u +3Y_d^\dagger Y_dY_d^\dagger Y_d\right\rbrace$, $T_{4\nu}\equiv \mathrm{Tr}\left\lbrace h_\nu^\dagger h_\nu h_\nu^\dagger h_\nu\right\rbrace$, and $T_{\nu e}\equiv \mathrm{Tr}\left( h_\nu^\dagger h_\nu Y_e^\dagger Y_e\right)$.
These are the most demanding equations in practice, since one needs the knowledge about the evolutions of all gauge and Yukawa couplings as input. Nevertheless, after having completed the program from Secs.~\ref{sec:RGE_gauge} and~\ref{sec:RGE_Yukawa}, it is no problem to also obtain the evolutions of the 4-scalar couplings $\lambda_i$.

Note that the corrections to all scalar couplings except for $\lambda_5$ are additive, while $\lambda_5$ only receives multiplicative corrections. This does not happen by accident, but rather there is a very deep reason for that: The Ma-model intrinsically involves the breaking of lepton number either by the heavy neutrino Majorana mass term or directly by the $\lambda_5$-term, cf.\ Sec.~\ref{sec:Ma-pure}, depending on which field, $N_k$ or $\eta$, lepton number is assigned to. However, even in the case of lepton number violation coming in directly by the heavy neutrino mass term, this breaking is not easily transmitted to the active neutrino sector. The decisive point is that the Dirac neutrino Yukawa coupling,  $\mathcal{L}'_{\rm Yuk,\nu} = - \overline{L} h_\nu \tilde \eta N_R + h.c.$, is lepton number \emph{conserving}. However, the diagram giving mass to the light neutrinos, cf.\ right panel of Fig.~\ref{fig:seesaws}, and the corresponding neutrino mass matrix, Eq.~\eqref{eq:numass-gen}, are proportional to $\lambda_5$, and they will transmit the violation of lepton number from the heavy to the light neutrino sector. Hence, in the case of $\lambda_5=0$, either the lepton number is exactly conserved (if $\eta$ carries lepton number) or violation is present in the model but will not translate into the light neutrino sector (if the $N_k$ carry lepton number). In any case, the form of Eq.~\eqref{eq:scal_run_e} is a protective limit of lepton number symmetry: If $\lambda_5=0$ holds at the input scale, it will always remain true, at any energy. This enforces all corrections to $\lambda_5=0$ to vanish identically, which is only possible if they are proportional to some power of $\lambda_5$. In addition, supposed that we start with very small couplings (close to zero) at the high-energy scale, one could hope for the parameters $\lambda_{1,2,3,4}$ to naturally grow to the relatively large values needed, while $\lambda_5$ remains very small. However, as we will see, at least for $\lambda_2$ such a magical behaviour is destroyed by the requirement of keeping vacuum stability. On the other hand, it is still true for $\lambda_{1,3,4,5}$.

We have also calculated the RGEs for the scalar masses in the Higgs potential:
\begin{subequations}
\begin{eqnarray}
\mathcal{D}m_1^2 &=& 6\lambda_1 m_1^2 +\left( 4\lambda_3+2\lambda_4\right)m_2^2 +m_1^2\left[ 2 T -\frac{3}{2}\left( g_1^2+3g_2^2\right) \right], \label{eq:mrun_a}\\
\mathcal{D}m_2^2 &=& 6\lambda_2 m_2^2 +\left( 4\lambda_3+2\lambda_4\right)m_1^2 + m_2^2\left[ 2 T_\nu -\frac{3}{2}\left( g_1^2+3g_2^2\right)\right] . \label{eq:mrun_b}
\end{eqnarray}
\end{subequations}
Note that the neutrino Yukawa coupling matrix only contributes to $m_2^2$, while all other Yukawa couplings contribute only to $m_1^2$. Furthermore, we also have to input the evolutions of $\lambda_{1,2,3,4}$ and of the gauge couplings $g_{1,2}$ into the equations for the scalar mass parameters.

\subsection{\label{sec:RGEs_EFT}Running in the effective theories}

Finally, we have computed the RGEs in the different effective theories. As explained, when integrating out the heavy parameters of the theory (chosen to be the heavy neutrino singlets), we switch from the full theory to the tower of effective theories, and the set of RGEs is altered compared to the FT. Hence, we need RGEs for the $d=5$ effective operators which generate a neutrino mass at low energies.

In the context of the THDM, such an operator is potentially split into four parts~\cite{Antusch:2001vn,Babu:1993qv}. However, as explained in Sec.~\ref{sec:Ma-EFT}, the $Z_2$ symmetry allows only two terms which we renormalize. The $\beta$-functions of the remaining two operators are:\footnote{This result is in agreement with the more general case computed in Ref.~\cite{Ibarra:2011gn}. However, if one wants to compare our results with the ones presented there, it is important to take into account the differences in the definitions of the coefficients in the scalar potential.}
\begin{subequations}
\begin{align}
\mathcal{D}\kappa^{(11)} &=-\frac{3}{2}\left\lbrace \kappa^{(11)}(Y_e^\dagger Y_e)+( Y_e^\dagger Y_e)^T\kappa^{(11)} \right\rbrace +2 T \kappa^{(11)} + 2\lambda_1\kappa^{(11)}+2\lambda_5\kappa^{(22)}-3g_2^2\kappa^{(11)}, \label{eq:effrun_a}\\[15pt]
\mathcal{D}\kappa^{(22)} &=\frac{1}{2}\left\lbrace \kappa^{(22)}(Y_e^\dagger Y_e)+(Y_e^\dagger Y_e)^T\kappa^{(22)} \right\rbrace + 2\lambda_2\kappa^{(22)}+2\lambda_5\kappa^{(11)}-3g_2^2\kappa^{(22)}. \label{eq:effrun_b}
\end{align}
\end{subequations}

This allows to write the general form of the RGEs throughout the different effective theories.

\section{\label{sec:Numerics}The numerical solution of the RGEs}

Finally, we present a numerical analysis of the equations presented in Sec.~\ref{sec:RGEs}. Since the whole set of RGEs is rather complicated, and a detailed numerical study of many example setups is beyond the scope of this paper, we chose to investigate one approximate case to illustrate the most generic features of the RGEs of the Ma-model. Note that this case is not more than an example, and even if we have a fair agreement with experiments, we would still need a more complete investigation in order to make solid predictions. We propose such a detailed numerical investigation for future work.

In our calculation, we make use of the following approximations:
\begin{itemize}

\item The top-quark Yukawa coupling being much larger than the other Yukawa couplings, we will assume that down-quark and charged lepton Yukawa matrices can be neglected compared to the up-quark Yukawa matrix, of which only the top-quark component will be considered: \mbox{$Y_d\,,\,Y_e\ll Y_u\sim {\rm diag} (0, 0, y_t)$.}

\item The down-quark and charged lepton Yukawa matrices will be neglected compared to the gauge couplings: $Y_d\,,\,Y_e\ll g_i,\,\, \text{where}\,\,i=1,2,3$.

\item We consider the charged lepton mass matrix to be diagonal. This simplification is motivated by the weak running of the charged lepton masses, and for the sake of an easy example one can accept the small errors introduced by it.

\end{itemize}

In addition to these approximations, we also make certain simplifying assumptions:
\begin{itemize}

\item As has been said previously, we assume that the inert scalar mass parameter ($m_2$) is below the masses of the heavy neutrinos ($M_{1,2,3}$), so that we can integrate these out while keeping the inert scalar as dynamical particle in the model. We take the physical mass to be $m_\eta \approx 200$ GeV at the $Z$-pole, which is in agreement with the upper bound derived from LHC measurements~\cite{Arhrib:2012ia}: $m_{\eta^\pm} \lesssim 200$ GeV.

\item We take the neutrino Yukawa coupling matrix $h_\nu$ to have real entries only, for simplicity.

\item We assume normal ordering for the light neutrino masses.

\item We assume that there is bimaximal mixing at the GUT (Grand Unified Theory) scale, $M_{\rm GUT}=10^{16}$ GeV.

\end{itemize}

Equipped with these approximations and assumptions, we will perform the numerical study using the \emph{top-down} approach, i.e.,~we will run the parameters from high-energy to low-energy, using input values at the GUT scale. As outlined in Sec.~\ref{sec:RGEs}, the set of RGEs will be solved step by step starting with the simplest non-coupled ones. The evolution of the gauge couplings and of the top-Yukawa coupling had already been presented in Fig.~\ref{fig:SM-coup}. These solutions are independent of the particular setting considered for the neutrino sector, and can hence taken to be universally true.\footnote{Note that this statement is only true as long as there is no flavour symmetry dictating the form of the charged lepton Yukawa coupling $Y_e$.} However, what we are mainly interested in is the neutrino sector and its interplay with the scalar sector, which will be investigated in the following.

\subsection{\label{sec:Numerics_couplings}The evolution of the couplings}

Following our previous approximations, in the FT, we only take neutrino Yukawa couplings and the top Yukawa coupling into account. For the neutrino Yukawa coupling matrix $h_\nu$, this implies a system of nine differential equations, one for each element of $h_\nu$.\footnote{Note that we take the couplings to be real, for simplicity. In the general case, however, one would instead have nine complex equations, i.e., eighteen real ones, minus the number of phases that can be absorbed by redefinitions of the fermion fields.} As far as the input values are concerned, we choose them such that we start from a bimaximal mixing scheme~\cite{Barger:1998ta} at the GUT scale. Although this mixing pattern is simply our assumption for the starting values at the input scale, it is nevertheless a well-motivated choice, as there are many flavour symmetries that tend to predict bimaximal mixing~(see, e.g., Refs.~\cite{Mohapatra:1998bp,Shafi:2000su,Ma:2001md,Kuchimanchi:2002yu,Altarelli:2009gn,Merlo:2011vc,Meloni:2011fx}). At the GUT scale, this means that we choose our Yukawa coupling matrix to be given by
\begin{equation}
 h_\nu^\text{input}(M_{\rm GUT})=h_\nu^\text{diag}(M_{\rm GUT})\,U^\dagger_\text{bimax},
 \label{eq:Yuk_input_1}
\end{equation}
where we have taken the input values $h_\nu^\text{diag}(M_{\rm GUT}) = {\rm diag} (0.245, 0.42, 0.1)$, and $U_\text{bimax}$ is the bimaximal mixing matrix,
\begin{equation}
 U_\text{bimax}=\begin{pmatrix}
 \frac{1}{\sqrt{2}} & \frac{1}{\sqrt{2}} & 0\\
 -\frac{1}{2} & \frac{1}{2} & \frac{1}{\sqrt{2}}\\
 \frac{1}{2} & -\frac{1}{2} & \frac{1}{\sqrt{2}}
 \end{pmatrix}.
 \label{eq:Yuk_input_2}
\end{equation}
Note that Eq.~\eqref{eq:Yuk_input_1} encodes the leptonic mixing correctly at the GUT scale, where we (arbitrarily) assume the charged lepton and heavy neutrino mass matrices to be diagonal. However, the running will in general introduce non-diagonal elements to all these matrices, which means that we have to make use of Eqs.~\eqref{eq:mixpar_a} to~\eqref{eq:mixpar_d} at lower scales. In particular, although it is perfectly justified to neglect the charged lepton Yukawa couplings in the FT at high scales, this is not valid anymore when we solve the equations for the effective operators. The reason for this is that there is a direct competition between $\lambda_5$ and $Y_e$, cf.\ Eq.~\eqref{eq:effrun_b}. Hence, in the effective theories, we also input the charged lepton masses in the RGE for $\kappa^{(22)}$. Accordingly, we have taken into account the running of $Y_e$ in ET$3$, ET$2$, and ET$1$. However, as this running is not expected to be very strong, we have simply used the low-energy experimental values of the charged lepton masses as input at the high scale, and indeed their values turned out to hardly change in the evolution.

\begin{figure}[t]
\centering
\includegraphics[width=10cm]{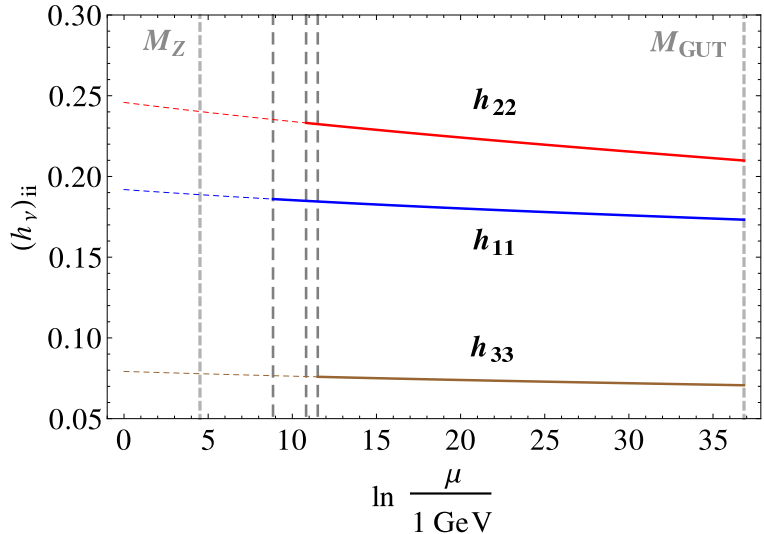}
\caption{\label{fig:couplings_run} Evolutions of the (diagonal) neutrino Yukawa couplings. The boundaries between the different theories, as well as the high- and low-energy scales, are indicated by vertical dashed lines.}
\end{figure}

As illustration for the running of the neutrino couplings, we have depicted the evolution of the diagonal entries of $h_\nu$ (not to be confused with $h_\nu^\text{diag}$) in Fig.~\ref{fig:couplings_run}. We could also have presented a plot for the non-diagonal entries, but we prefer to focus on the actual masses and mixing angles, to be presented below, as these quantities are much easier to interpret physically. Note that, since some of the Yukawa couplings fail to be defined after certain heavy neutrinos have been integrated out (cf.\ Sec.~\ref{sec:Ma-tching}), their extrapolated evolution is depicted by dashed lines in Fig.~\ref{fig:couplings_run}. As we can see from the figure, their running is relatively moderate. However, we have to take into account that all nine entries of $h_\nu$ will run, and that the structure of the light neutrino mass matrix, Eq.~\eqref{eq:numass-gen}, is also strongly influenced by the running of the heavy neutrino and scalar parameters.

\begin{figure}[t]
\centering
\includegraphics[width=8.0cm]{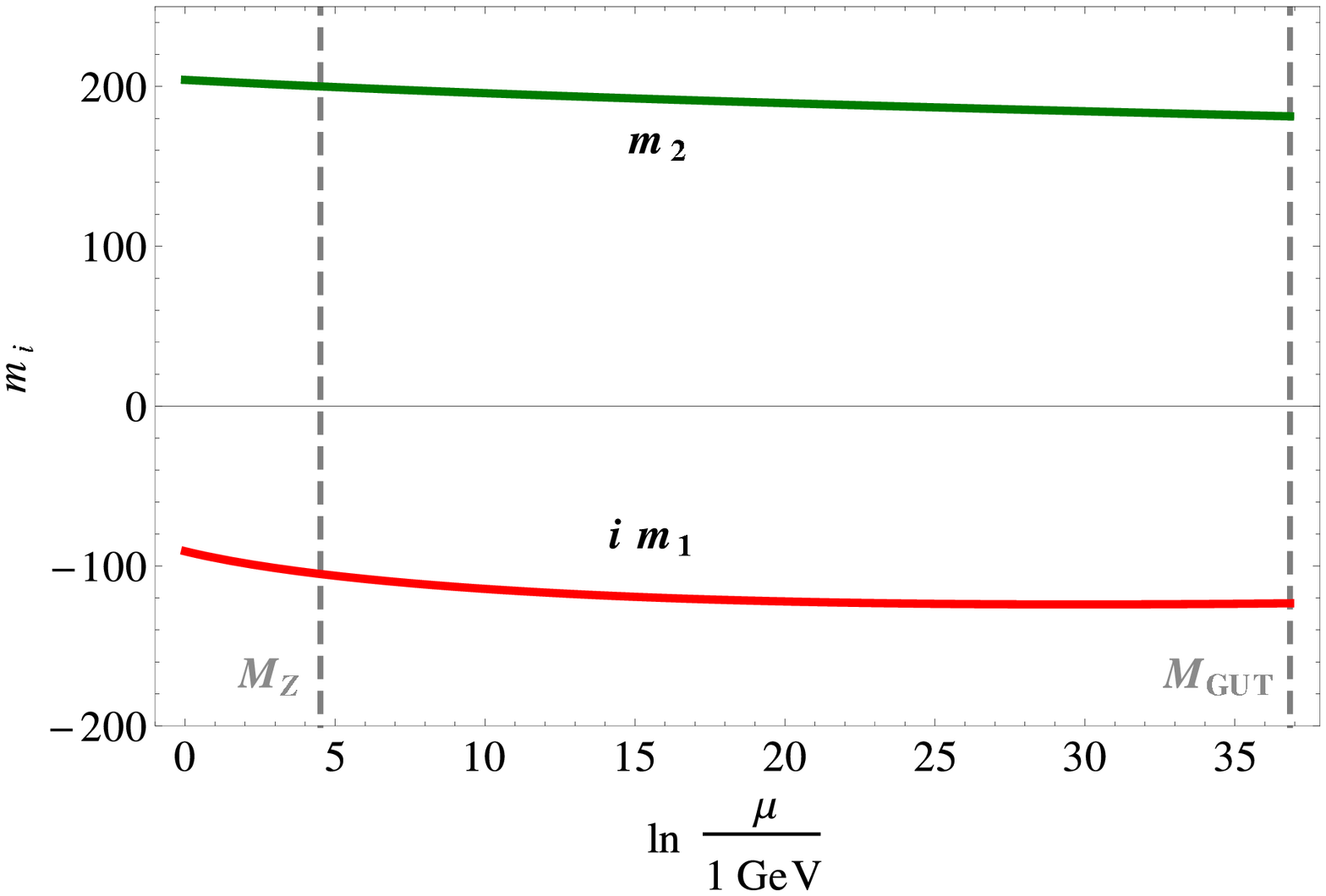}
\includegraphics[width=7.8cm]{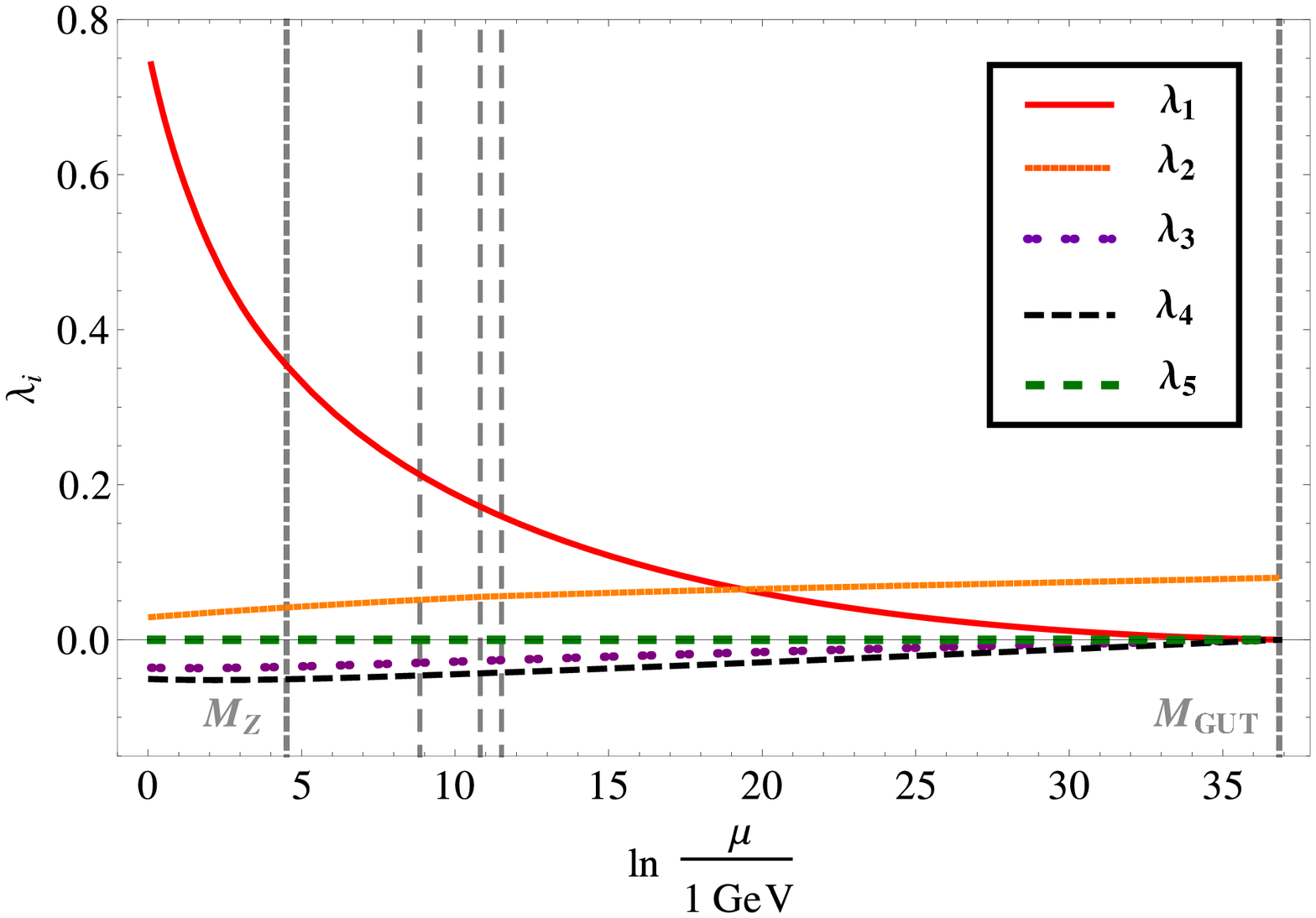}
\caption{\label{fig:scalpar_run} The running of the scalar parameters. Note that $m_1^2$ has to be negative, in order to guarantee a VEV for the SM-like Higgs, cf.\ Sec.~\ref{sec:Ma-pure}.}
\end{figure}

\begin{figure}[t]
\centering
\includegraphics[width=10cm]{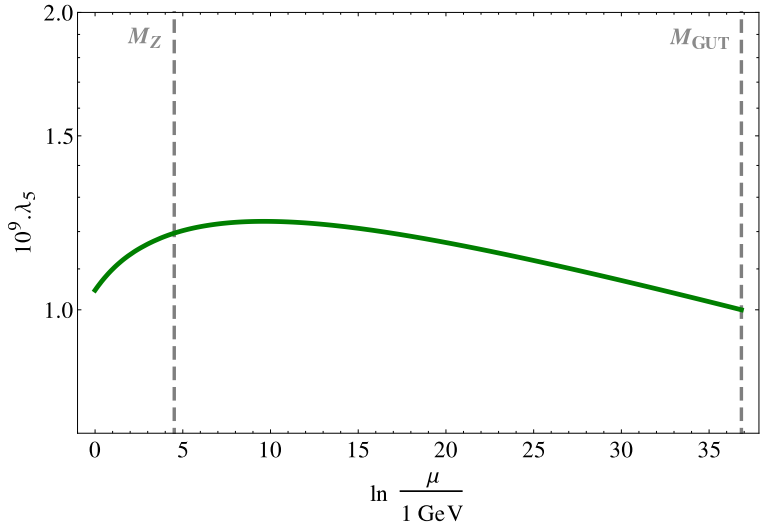}
\caption{\label{fig:lambda5_run} Detailed evolution of the lepton number violating coupling $\lambda_5$.}
\end{figure}

The RGEs for the scalar couplings, Eqs.~\eqref{eq:scal_run_a} to~\eqref{eq:scal_run_e} and~\eqref{eq:mrun_a} to~\eqref{eq:mrun_b}, make use of those for $g_i$, $h_\nu$, and $Y_e$. All these equations are interdependent, and therefore we have to solve the whole system at once. The initial values we choose are:
\begin{equation}
\lambda_{1,3,4,5} (M_{\rm GUT}) = 10^{-9}\ \ \ {\rm and}\ \ \ \lambda_2(M_{\rm GUT})=0.08,
 \label{lambda_input}
\end{equation}
where $M_{\rm GUT}$ again denotes the GUT scale. These initial values were chosen such that three requirements are fulfilled:
\begin{enumerate}

\item The correct relation between $m_1$ and the coupling $\lambda_1$ (obtained after EWSB) holds at the electroweak scale: $v^2 = \frac{- m_1^2}{\lambda_1}$.

\item Vacuum stability is ensured.

\item The masses of the particles originating from the inert doublet remain consistent with collider constraints (i.e., $m_{\eta^\pm} \lesssim 200$ GeV~\cite{Arhrib:2012ia}). 

\end{enumerate}
Although it would seem natural to choose the same input value for $\lambda_2$ as for all the other couplings there arises a problem, as already mentioned in Sec.~\ref{sec:RGE_scalars}: Since $\lambda_2$ decreases with decreasing energy, we have to choose its input value large enough not to have $\lambda_2<0$ at some point, i.e., not to break vacuum stability. This behaviour stems from the form of Eq.~\eqref{eq:scal_run_b}, whose right-hand side has a strong tendency to be positive, as can be seen easily by realizing that the term $4\lambda_2 T_\nu $ will be dominant as long as we have sizable Yukawa couplings $h_\nu$ which are nevertheless in the perturbative range. However, one might be able to improve on this point by studying a more general scenario. Still, the parameter choice in Eq.~\eqref{lambda_input} is very interesting because, as expected, all couplings receive strong corrections with the exception of $\lambda_5$, which remains of $\mathcal{O}(10^{-9})$. This behaviour, together with the (moderate) running of the mass parameters $m_{1,2}$, is presented in Fig.~\ref{fig:scalpar_run}. In addition to that, we have plotted the running of $\lambda_5$, drawn to a larger scale, in Fig.~\ref{fig:lambda5_run}. Indeed, $\lambda_5$ does run, but its running is weak due to the fact that $\lambda_5$ only receives multiplicative corrections, while the other couplings all have additive contributions.

\subsection{\label{sec:Numerics_masses}Masses and mixings}

\begin{figure}[t]
\centering
\includegraphics[width=10cm]{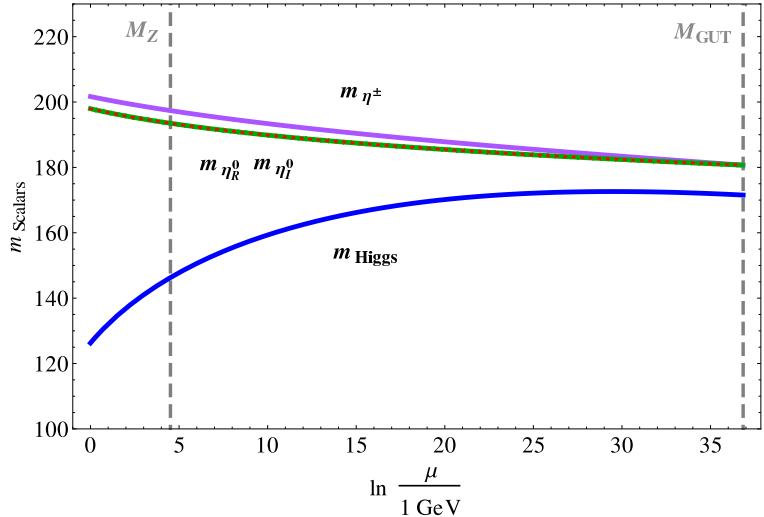}
\caption{\label{fig:scalmas_run} Evolution of the physical scalar masses. }
\end{figure}

Let us now turn to masses and mixing angles. First, we want discuss the evolution of the physical scalar masses, depicted in Fig.~\ref{fig:scalmas_run}. This plot is obtained by simply inserting the runnings of all scalar parameters, cf.\ Fig.~\ref{fig:scalpar_run}, into the explicit expressions for the masses, Eqs.~\eqref{eq:scalar_masses}. Due to the absence of a VEV for $\eta$, the coupling $\lambda_2$ does \emph{not} contribute to the scalar masses directly, which is why indeed the running of the inert scalar masses is mainly driven by the additive corrections to $\lambda_3$. The splitting between the electrically charged and the electrically neutral components of $\eta$ arises due to the evolution of $\lambda_4$, while the splitting between the real and imaginary parts of $\eta$ is only proportional to the small coupling $\lambda_5$, and hence it is not visible in the plot. Nevertheless, the smallness of the mass different between $\eta^0_R$ and $\eta^0_I$ is decisive for the correct neutrino masses, and a small splitting is enforced by the approximate lepton number conservation in the model. 

Note that we predict a relatively large value for the SM-like Higgs mass at low energies, $m_h \approx 140$~GeV. This is mainly due to the strong corrections to the parameter $\lambda_1$, cf.\ right panel of Fig.~\ref{fig:scalpar_run}, which arise from a complicated interplay of all other parameters. However, we are still very close to the range that is indicated by the newest LHC data from ATLAS~\cite{ATLAS:2011aa,ATLAS:2012ad} and CMS~\cite{Chatrchyan:2012tw,Chatrchyan:2012tx}, so that we can hope to obtain similar (or even better) predictions by taking into account, e.g., some more Yukawa couplings.

\begin{figure}[t]
\centering
\includegraphics[width=7.5cm]{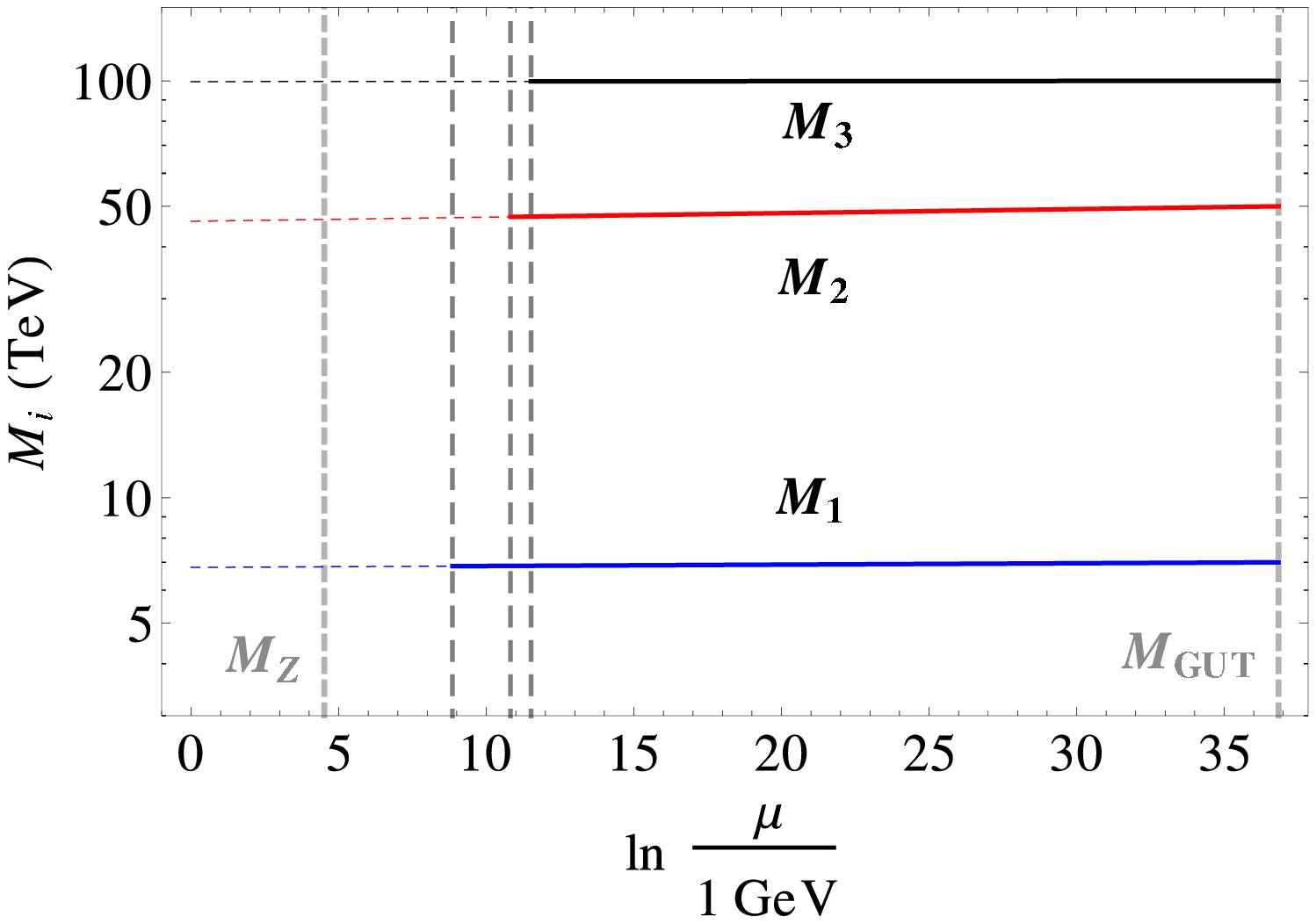}
\includegraphics[width=8.1cm]{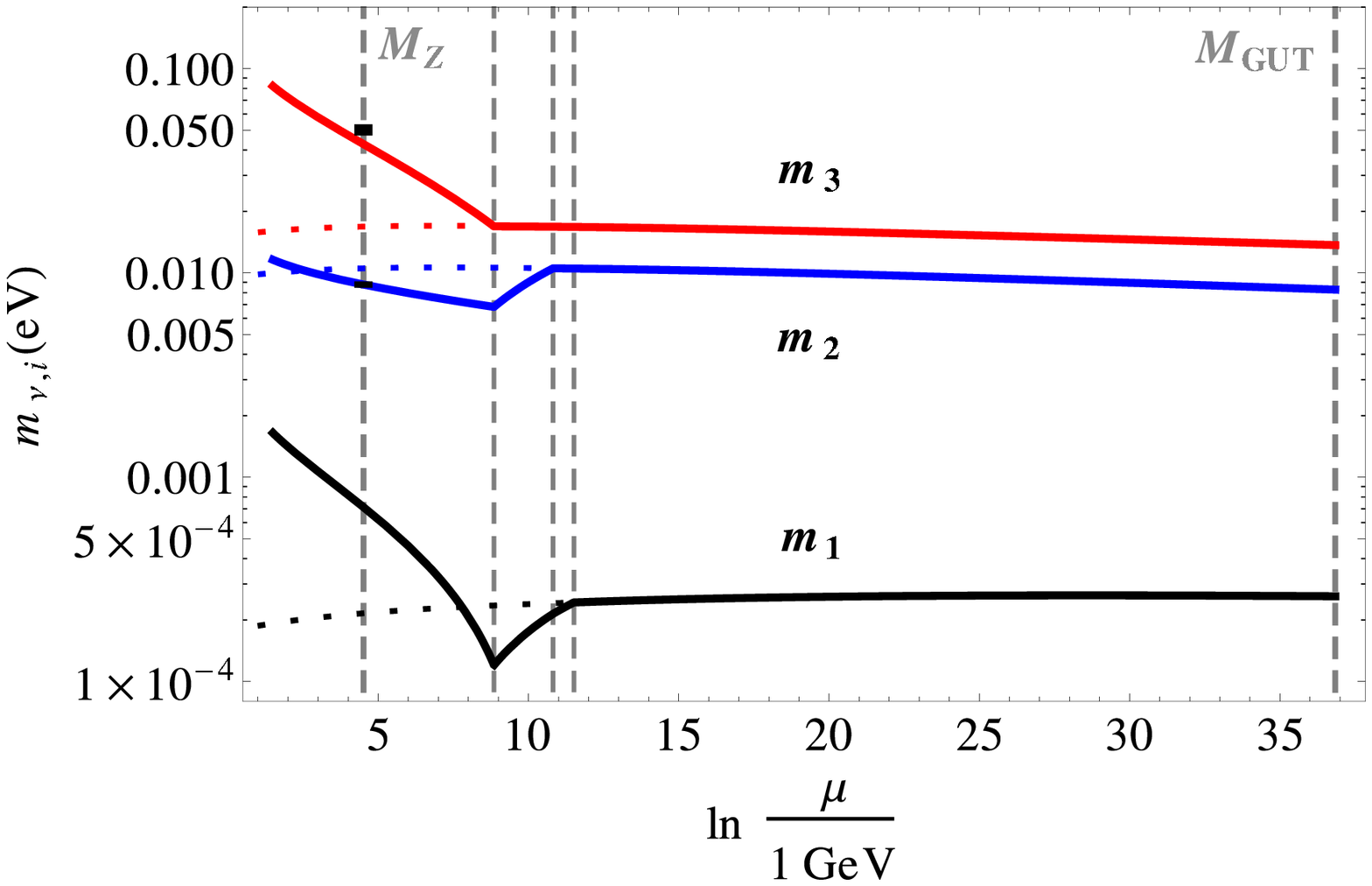}
\caption{\label{fig:neutrino_run} The running of the heavy and light neutrino masses.}
\end{figure}

Next we turn to the heavy neutrino masses, i.e., the solutions of Eq.~\eqref{eq:N-evol}. The input values at the GUT were chosen such that we get consistent light neutrino parameters. The configuration we use is:
\begin{equation}
(M_1,M_2,M_3)=(7,50,100)~{\rm TeV},
\end{equation}
at the scale $\mu=M_{\rm GUT}$. The corresponding evolution is depicted in the left panel of Fig.~\ref{fig:neutrino_run}. Again, these masses are not really defined in the model below the scales where the corresponding heavy neutrinos have been integrated out, and hence again the use of dashed lines in the plot. However, of course these particles still ``exist'', in the sense that they have visible contributions to low-energy physics, in particular the existence of non-zero light neutrino masses. It is only that these heavy neutrinos are kinematically decoupled. Note that the corrections to these masses actually seem quite moderate in the plot. One has to keep in mind, though, that the absolute corrections to the heavy neutrino masses are indeed quite large (several hundreds of GeV), and it is just the high scale of their masses and the multiplicative nature of their corrections, cf.\ Eq.~\eqref{eq:N-evol}, which render their relative correction to be small.

Finally, we want to discuss the light neutrino sector, which is probably the most important and interesting result in our simplified scenario. The evolution of the neutrino mass matrix can be determined by inserting the results of the relevant RGEs into Eq.~\eqref{eq:numass-gen}. Hence, we actually do not choose directly any input values for the light neutrino masses at the GUT-scale, but we rather use the input values of the neutrino Yukawa and scalar couplings, as well as the ones of the heavy neutrino masses, which are all explicitly specified at the input scale. We can then determine the evolution of the neutrino mass matrix $\mathcal{M}_\nu (\mu)$, as a function of the energy scale $\mu$. By calculating the eigenvalues of $\mathcal{M}_\nu$, we can determine the evolution of the light neutrino mass eigenvalues, depicted on the right panel of Fig.~\ref{fig:neutrino_run}. Note that we have chosen to work with a normal neutrino mass ordering, as explained above. We can see from the plots that the running of the light neutrino masses, and in particular the threshold effects, are relatively strong. This was to be expected, as there are many couplings involved in the physical neutrino mass matrix whose effects could potentially enhance each other. By taking a closer look at the plot, we can also recognize the seesaw structure: $N_3$ is the heavy neutrino with the largest mass, and it gets integrated out at the boundary between FT and ET$3$. Since, however, the inverse of the heavy neutrino matrix leads, by the seesaw structure, to the strongest suppression of the light neutrino masses, one can see the threshold effect for the lightest neutrino mass $m_1$ at exactly that scale, while the other masses are hardly influenced. Similarly, integrating out $N_2$ results into a strong threshold effect for $m_2$, and integrating out $N_1$ strongly influences $m_3$.

Our goal is to reproduce the neutrino mass eigenvalues at low energies. However, as explained in the introduction, currently we still have no knowledge of the absolute neutrino mass scale, apart from upper bounds. Hence, we compare our low-energy predictions to the experimental values by ``fixing'' the light neutrino mass scale by the value obtained for $m_1$ at the $Z$-pole. We then take the values measured for the neutrino mass-square differences~\cite{Schwetz:2011qt,Schwetz:2011zk} to determine the experimentally preferred $3\sigma$ ranges for $m_{2,3}$, which are indicated by the black boxes in Fig.~\ref{fig:neutrino_run}, right panel. Note that the experimental ranges are very narrow, and we only hit the corresponding range with $m_2$, while our prediction for $m_3$ is close to but still below the experimentally preferred range. On the other hand, as already explained, we are looking at a simplified scenario here, so we cannot expect to have a perfect match to the experimental values. Furthermore, we often measure the mass square differences at even lower energies (e.g.\ $\mathcal{O}$(MeV) for reactor neutrinos or $\mathcal{O}$(GeV) for atmospheric neutrinos~\cite{Barger:2003qi}), and our figure shows considerable running even below that scale. Apparently, it would be worth to perform a detailed investigation of this region in the Ma-model.

\begin{figure}[t]
\centering
\includegraphics[width=13cm]{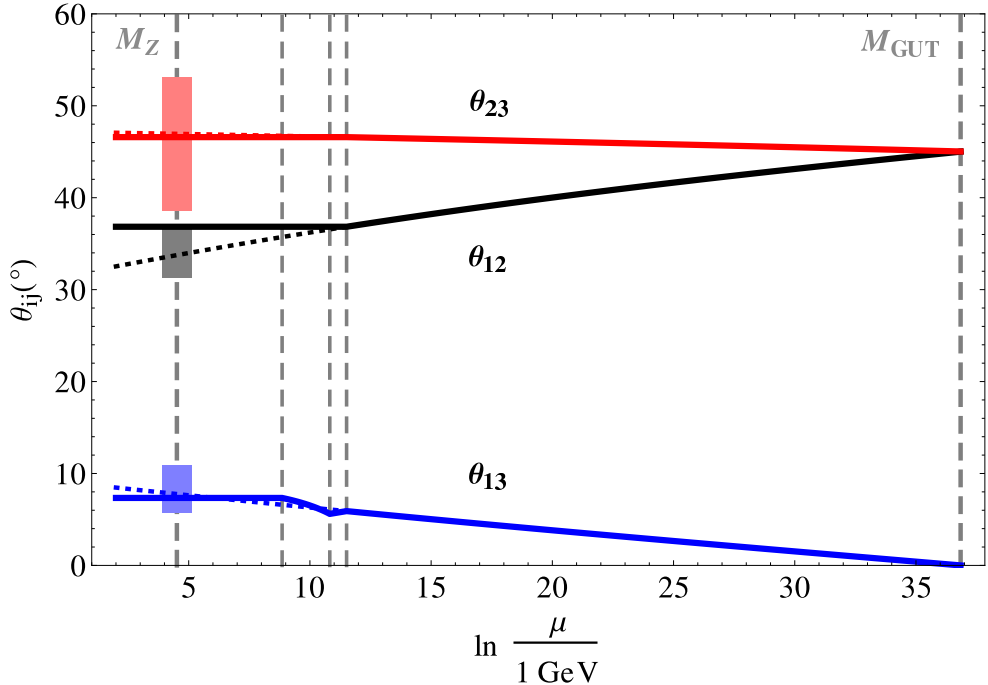}
\caption{\label{fig:mixings_run} Evolution of the leptonic mixing angles.}
\end{figure}

Finally, the evolution of the three neutrino mixing angles is depicted in Fig.~\ref{fig:mixings_run}. It is this picture that illuminates the potential of our considerations most clearly: In the figure, the experimental 3$\sigma$ ranges for the mixing angles $\theta_{12}$ (black) and $\theta_{23}$ (red)~\cite{Schwetz:2011qt,Schwetz:2011zk}, as well as the newly measured $\theta_{13}$ (blue)~\cite{An:2012eh} are indicated by the light colored boxes at the low-energy scale. This plot exhibits a remarkable behaviour of the Ma-model: We have started off with a bimaximal mixing pattern at the GUT scale, which could easily be motivated by a flavour symmetry.\footnote{Note that the simplest flavour symmetries have a certain preference for predicting maximal and zero mixing.} However, this mixing pattern is not compatible with low-energy data. But what happens in our setting is that the evolution of the leptonic mass matrices is such that $\theta_{12}$ decreases and $\theta_{13}$ increases with decreasing energy, while $\theta_{23}$ remains practically constant. Hence, we have a remarkably good match of our predictions with the low-energy experimental ranges. In particular, the corrections have a tendency to lead to a relatively large value for $\theta_{13}$, which is in perfect agreement with the very recent measurements by Daya Bay and RENO. In that respect, our result is timely and suggests a more detailed study of the evolution of the parameters in the Ma-model, which we leave for future work.

Note that, of course, our low-energy predictions strongly depend on the input values chosen at the GUT-scale. Had we chosen other parameters, we would have obtained different predictions. It had not been guaranteed that there is a certain tendency in the Ma-model in our setting not only to predict the correct mixing angles but also to be in agreement with the ranges or bounds for many other observables, but it actually turns out to be the case, at least in our example scenario. This is non-trivial, and it was not a priori clear that this would happen.

Furthermore, there is an interesting test of the high-energy sector: One could apply different flavour symmetries to predict a set of mixing patterns at the GUT scale (or at some other high scale) and evolve the neutrino observables down to low energies. Then, depending on the high-energy prediction, certain models would be in agreement with low-energy phenomenology, while others would violate certain conditions. In addition, in this setting, there are also alternative observables as, e.g., the inert scalar masses. This could even be extended to include, e.g., heavy neutrino or inert doublet Dark Matter or LFV processes. Hence, one could use the Ma-model as a tool to tell apart successful from failing symmetries: If a certain flavour symmetry predicts a high-energy pattern that does not yield low-energy agreement, we can discard it in the context of the Ma-model. This is particularly interesting since even relatively simple flavour symmetries, which predict a simple mixing pattern at high energies, could still be compatible with low-energy data. There is a rich variety of similarly interesting applications of our RGEs which could be studied in the future.

\section{\label{sec:Conclusions}Conclusions and outlook}

In this paper, we have investigated the 1-loop renormalization of Ma's scotogenic model, which is one of the easiest extensions of the Standard Model leading to a non-zero neutrino mass at 1-loop level. After having introduced the model itself, we have discussed how the different effective theories arise when integrating out the heavy neutrino mass eigenstates one by one. This results into three different effective theories at energies below the full theory. When calculating the renormalization, we need to match all operators that contribute to the neutrino mass at 1-loop level whenever a ``new'' effective theory arises. We have calculated the renormalization group equations for the full theory as well as for all three effective theories, which led to a very general expression for the light neutrino mass matrix in terms of various couplings present in the model. Furthermore, we have performed a numerical study of a simplified scenario, in order to give an illustration of the renormalization effects that arise in the model. As to be expected for a loop mass, the running and in particular the threshold effects are relatively strong: While the leptonic mixing angles exhibit considerable running at high energies, the running of the light neutrino mass eigenvalues originates mainly from threshold effects. We have found a particularly nice setting that suggests interesting connections to flavour symmetries: We start off with bimaximal mixing pattern at the grand unification scale, which evolves towards low energies into a pattern that matches very well the experimentally allowed ranges for the mixing angles. In addition, also the neutrino mass eigenvalues and the Higgs masses are at least close to the experimentally interesting regions.

As we have seen, the renormalization of radiative neutrino mass models can involve a very rich and interesting phenomenology. While we have studied the case of the scotogenic model, it could be similarly beneficial to compute renormalization group equations for other models with a loop-mass for neutrinos, such as the Zee-Babu model. In all these models, detailed phenomenological studies could be performed, in order to test the validity of the models thoroughly. It could be that the running is, e.g., too strong in certain cases, thereby tending to disagree with certain experimental bounds, or it could be that the Higgs sector in some of those models have severe problems with vacuum stability. In terms of neutrino physics, and in particular in the light of the recent measurements of $\theta_{13}$, maybe one of the most interesting points to study is the combination of flavour symmetries with the known radiative neutrino mass models. We hope that many interesting such studies will appear in the future.

\section*{Acknowledgements}

We would like to thank T.~Ohlsson and M.~A.~Schmidt for useful discussions, and we are particularly grateful to J.~Bergstr\"om for carefully reading the manuscript and giving valuable comments. The work of AM is supported by the G\"oran Gustafsson foundation.

\bibliographystyle{./apsrev}
\bibliography{./Rad-Running}

\end{document}